\documentclass{aa}
\usepackage{graphics}
\begin{document}
\def\aa{$\mathrm{COMA}$}
\def\bb{$\mathrm{A194}$}
\def\cc{$\mathrm{A539}$}
\def\dd{$\mathrm{A3381}$}
\def\ee{$\mathrm{S639}$}
\def\ff{$\mathrm{S753}$}
\def\gg{$\mathrm{HYDRA}$}
\def\hh{$\mathrm{PERSEUS}$}
\def\ii{$\mathrm{PISCES}$}
\def\jj{$\mathrm{J8}$}
\def\hyp{$\mathcal{H}_1$}
\def\hys{$\mathcal{H}_{2}$}
\def\hyt{$\mathcal{H}_3$}
\def\hyq{$\mathcal{H}_4$}
\def\hytp{$\mathcal{H'}_3$}
\def\hyqp{$\mathcal{H'}_4$}
\def\hyqu{$\mathcal{H}_5$}
\def\sp{$\mathcal{S}_1$}
\def\ss{$\mathcal{S}_{2}$}
\def\op{$\mathcal{R}_1$}
\def\os{$\mathcal{R}_{2}$}
\def\ot{$\mathcal{R}_3$}
\def\oq{$\mathcal{R}_4$}
\def\oqu{$\mathcal{R}_5$}
\def\osx{$\mathcal{R}_6$}
\def\ose{$\mathcal{R}_7$}
\def\ep{$\mathcal{E}_1$}
\def\es{$\mathcal{E}_2$}
\def\flr{$\mathrm{OLS}_{\log r_{\mathrm{e}}}$ }
\def\fbr{$\mathrm{OLS}_{\mu_{\mathrm{e}}}$ }
\def\fls{$\mathrm{OLS}_{\log \sigma_0}$ }
\def\for{$\mathrm{ORLS}$ }
\def\erlr{$\mathrm{MIST}_{\log r_{\mathrm{e}}}$ }
\def\erbr{$\mathrm{MIST}_{\mu_{\mathrm{e}}}$ }
\def\erls{$\mathrm{MIST}_{\log \sigma_0}$ }
\def\erbi{$\mathrm{MIST}_{\mathrm{BIS}}$ }
\def\lr{$\log r_{\mathrm{e}}$}
\def\br{$\mu_{\mathrm{e}}$}
\def\ie{$< \! I_{\mathrm{e}} \! >$}
\def\ls{$\log \sigma_0$}
\def\xxi{$\left\{ X_i \right\}$}
\def\yi{$\left\{ Y_i \right\}$}
\def\fmi{$ \left\{ {\Phi_{\mathrm{m}}}_i \right\} $} 
\def\fsi{$ \left\{ {\Phi_{\mathrm{s}}}_i \right\} $}
\def\smr{rms}
\def\fri{$\varepsilon$}

   \thesaurus{03     %
              (11.06.2;  
               03.13.6;  
               11.03.1;  
              }
   \title{Fitting planes to early-type galaxies: 
      MIST for the determination of the Fundamental Plane}
 
   \subtitle{}

   \author{F. La Barbera
          \inst{1}
          \and
           G. Busarello \inst{2}
          \and
           M. Capaccioli \inst{1,2}
           }

   \offprints{F. La Barbera}
   \institute{Universit\`a Federico II, Napoli, Italy \\
              email: labarber@na.astro.it
         \and
              Osservatorio Astronomico di Capodimonte, via Moiariello 
              16, I-81131 Napoli- Italy }
              \date{Received 30 June 2000 / accepted 6 September 2000
              }

\authorrunning{La Barbera et al.}
 \titlerunning{Fitting the Fundamental Plane}

   \maketitle

   \begin{abstract}
The present study deals with the problem of deriving the coefficients of
the fundamental plane (FP) of early-type galaxies.

We introduce statistical models of the FP and relative fitting methods:
the MIST ({\it Measurement errors and Intrinsic Scatter Three
dimensional}) fits\footnote{The FORTRAN codes of the MIST algorithms
are available on request to the authors.}. The MIST fits account for 
the measurement errors on the variables and their correlations as well as 
for the intrinsic scatter. We show that the lack of a model of the intrinsic 
scatter of the FP is the origin of the systematic differences between the 
various fitting methods.

We also address the problem of estimating the uncertainties of the FP
coefficients and determine a simple relation between the sample size and
the expected accuracy of the coefficients.

The present study leads to define a `minimum sample size' for a correct
estimate of the uncertainties. For $N \widetilde{<}30$, both theoretical
formulae and re-sampling techniques, like the bootstrap, do not give
reliable estimates.

The question of the 'universality' of the FP is addressed by applying the
MIST fits to ten samples of cluster galaxies. The FP slopes are actually
consistent between the different samples, but, due to the large
uncertainties, they could also hide significant systematic differences.

The feasibility of the measurement of the possible variations of the FP
slopes as a function of redshift is also proved.

  \end{abstract}

%

\section{Introduction.}

The fundamental plane (FP) is a bivariate relation between observed
global properties of early-type (E) galaxies like the effective radius
$r_e$, the mean surface brightness $\mu_e$, and the central velocity
dispersion $\sigma_0$ (e.g. Dressler et al. \cite{dlb87}, 
Djorgovski \& Davis \cite{djd87}) of the 
form\footnote{This representation of the FP is adopted
throughout the paper.}:
\begin{equation}
 \log r_\mathrm{e} = a \log \sigma_0 + b \mu_\mathrm{e} + c\ .
\label{fpeq}
\end{equation}
The main features of the FP are its small scatter ($\sim 10\%$ in
$r_e$, of the same order of the measurement errors) and the
`tilt' of its slopes with respect to the prediction of
the virial theorem for a family of homologous systems with 
constant $M/L$ ratio. 

It is generally believed that the `tilt' carries information on the
nature of Es, although no convincing interpretation has been found
that can be reconciled with the small scatter (see e.g. Ciotti et al.
\cite{clr96}, Renzini \& Ciotti \cite{rec93}). It is now clear that
part of the `tilt' is due to the intrinsic non-homology, both
structural and dynamical, of the E family (see e.g. Busarello et al.
\cite{bcc97}, Graham \& Colless \cite{grc96}, Prugniel \& Simien
\cite{prs97} and references therein). Stellar population effects (e.g.
systematic differences in age/metallicity and interstellar matter
content) should account for the remaining tilt (see e.g. Mobasher et
al. \cite{mga99}). In that case, the tilt should decrease when moving
to near-infrared wavelengths, where these effects are minimized: some
evidences exist that this is actually the case (e.g. Pahre et al.
\cite{pcd98}, Scodeggio et al. \cite{sgb98} and references therein).
The wavelength dependence of the FP enters also in the comparisons
between FP determinations at different redshifts, where different
rest-frame wavebands are sampled.

The FP has been soon recognized as a precise tool for distance
determinations (e.g. Dressler et al. \cite{dlb87}, Hudson et al.
\cite{hls97}) and, more recently, for constraining cosmological
parameters and for studying galaxy evolution (van Dokkum et al.
\cite{vdf96}, Bender et al. \cite{bsz98}, J\o rgensen et al.
\cite{jor99}, Kelson et al. \cite{kivd00}, and references therein).

All the above applications are based on comparisons between
different determinations of the FP: e.g. different samples, different
wavebands, and different redshifts. The method 
used to derive the FP coefficients and the relative uncertainties 
play therefore a central r\^ole for a proper use of the FP.

The determination of the FP requires considerable observational and
data analysis efforts (see e.g. Ziegler \& Bender \cite{zib97} and Ziegler 
et al. \cite{zsb99}), it is thus desirable that
a similar effort be made for the derivation of the coefficients of the FP 
and for an accurate estimate of the relative uncertainties.

There is still no agreement in the scientific community about the
fitting method to adopt. This is partly due to the numerous
applications for which the FP is derived (see above), but it also
originates from two important points: (1) the measurement errors on
the FP variables are comparable (see e.g. Smith et al. \cite{slh97},
hereafter SLH97) and correlated (see J\o rgensen et al. \cite{jfk95a},
hereafter JFK95a), (2) the scatter around the plane is not completely
accounted for by the measurement uncertainties but has also an
intrinsic origin (see e.g. J\o rgensen et al. \cite{jfk96}, hereafter
JFK96).

In  the present work we address the problems relative to:
(I) the choice of the procedure to derive the FP coefficients (Sect.~2)
and (II) techniques to estimate the corresponding uncertainties
(Sect.~3).

Starting from the bi-dimensional models introduced by Akritas
\& Bershady (\cite{akb96}, hereafter AkB96), we propose in Sect.~2.1 
statistical models of the FP that account for the various sources of 
scatter on the variables and derive the relative fitting procedures: 
the MIST ({\it Measurement errors and Intrinsic Scatter Three dimensional}) 
fits.

We then discuss in Sect.~2.2 the capability of the fitting procedures to
derive the coefficients of the FP.

Section 3 deals with the uncertainties of the coefficients of the FP.
The simulation algorithm used for the present analysis
is described in Sect.~3.1. In
Sect.~3.2 the different methods to estimate the uncertainties are
analyzed and compared to the results of the simulations.

The results and the techniques developed in Sect.~2, 3, are applied
in Sect.~4 to the comparison of the FPs of different
clusters of galaxies.

\section{Deriving the FP.}

The choice of the best fitting procedure in the study of astronomical
data is generally a not trivial question: the appropriate method
should be always suggested by the scientific problem to be analyzed 
(see the discussion in Isobe et al. \cite{ifa90}, hereafter IFA90).

The derivation of the FP is usually based on least squares methods: the
ordinary least squares (OLS) fits (e.g. Lucey et al. \cite{lbe91}, 
Kj\ae rgaard et al. \cite{kjm93}, Hudson et al. \cite{hls97}), in which 
the root mean square (\smr ) of residuals relative to one of the variables 
is minimized, the orthogonal least squares (ORLS) fit (e.g. Busarello 
et al. \cite{bcc97}), in which it is minimized the \smr~
of residuals perpendicular to the plane, and other methods derived from
the previous ones, as the bisector fit (e.g. Graham \& Colless \cite{grc96}) 
and the arithmetic mean of the OLS coefficients (e.g. Faber et al. 
\cite{fdd87}). To reduce the effect of
outliers, the  fits are often performed by robust procedures, in which
the sum of absolute residuals is minimized (see JFK96).

Fitting methods based on multivariate analysis techniques, like the
principal component analysis (e.g. Bender et al. \cite{bbf92}), have been
sometimes adopted. By their nature, however, these methods are not suitable
for the determination of best-fit coefficients, so that they
will be not considered in the following.

The most natural interpretation of the FP is in terms of a 'mean relation'
between global quantities of E galaxies with respect to which they scatter
in the space of the observed properties. Drawing hint from AkB96, we will
take this interpretation into mathematical terms, by introducing a
statistical model for the FP (Sect.~2.1). The corresponding fitting
procedures will be also derived.

We will then discuss (Sect.~2.2) the problem of recovering the mean
relation and illustrate the origin of the dependence of the FP
coefficients on the fitting method.

\subsection{A statistical model: the MIST fits.}
Let us introduce three random variables   
$\left\{ X_i \right\}_{i=1,2,3}$, describing the distribution of 
global quantities of E galaxies.
We will assume  that  \xxi~  verify the following identity:
\begin{equation}
 X_3 = \alpha X_1 + \beta X_2 + \gamma 
\label{relmed}
\end{equation}
where $\alpha$ and $\beta$ are the slopes and $\gamma$ is the zero 
point of the mean relation. 

In the following, we will use capitals to indicate random variables (RV)
and the corresponding small letters for their outputs. Given two RVs $A$
and $B$, we will indicate with $\mathrm{C}(A,B)$ their covariance, with
$\mathrm{V}(A)$ and $\mathrm{V}(B)$ their variances and with
$\mathrm{E}(A)$ and $\mathrm{E}(B)$ their expected values. The estimators
of a given quantity (i.e. the RVs used to approximate the quantity we look
for) will be marked with a caret (e.g. $\hat{\alpha}$, $\hat{\beta}$ and
$\hat{\gamma}$). An estimator will be said `unbiased' if its expected value
coincides with the quantity to be estimated.
 
In modeling the scatter around the plane defined by Eq.~(\ref{relmed}),
we have not only to consider measurement errors on the variables but
also an intrinsic dispersion.

To this aim, let us introduce two sets of RVs, 
$ \left\{ {\Phi_\mathrm{s}}_i \right\}_{i=1,2,3}$ and
$ \left\{ {\Phi_\mathrm{m}}_i \right\}_{i=1,2,3}$, with zero expected
values, that describe respectively the intrinsic dispersion and the
scatter due to measurement errors, and let us consider the following
relations:
\begin{equation}
 Y_i = X_i + {\Phi_\mathrm{m}}_i + {\Phi_\mathrm{s}}_i
\label{relsca}
\end{equation}
where $\left\{ Y_i \right\}_{i=1,2,3}$ are the RVs that describe
the distribution  of the observed quantities of E galaxies
(e.g. $Y_1 =${$\log r_\mathrm{e}$}, 
 $Y_2 =${ $\mu_\mathrm{e}$},
 $Y_3 =${ $\log \sigma_0$}).

The problem is then to determine unbiased estimators 
of $\alpha$, $\beta$  and $\gamma$. 

From Eq.~(\ref{relmed}), one obtains the following identities:
\begin{eqnarray}
 \mathrm{E}( X_3 ) & = & \alpha \cdot \mathrm{E}( X_1 ) + \beta \cdot
         \mathrm{E}( X_2 ) + \gamma \label{globa} \\ 
 \mathrm{C}( X_2 , X_3 ) & = & \alpha \cdot \mathrm{C}( X_1 , X_2 )
         + \beta \cdot \mathrm{C}( X_2 , X_2 ) =  \nonumber \\
           & = &  \alpha \cdot \mathrm{C}( X_1 , X_2 ) +
            \beta \cdot \mathrm{V}( X_2 )
        \label{globb} \\
 \mathrm{C}( X_1 , X_3 ) & = & \alpha \cdot \mathrm{C}( X_1 , X_1 )        
 + \beta \cdot \mathrm{C}( X_1 , X_2 )         =  \nonumber \\      
 & = & \alpha \cdot \mathrm{V}( X_1 ) + \beta \cdot \mathrm{C}( X_1 , X_2 ) 
 \label{globc}
\end{eqnarray}
where \begin{math} \mathrm{C}( X_i , X_j )  \end{math} 
and \begin{math} \mathrm{V}( X_k) \end{math} 
are the components of the covariance matrix (CM) of 
$\left\{ X_i \right\}$.
Assuming that \xxi, \fmi~ and \fsi~ are mutually not correlated 
(hereafter `hypothesis \hyp') and making use of Eqs.~(\ref{relmed}) 
and (\ref{relsca}), we can express the first and second order moments 
of \xxi~ as (see App. A for a straight demonstration and compare 
AkB96):
\begin{eqnarray}
 \mathrm{E}( Y_i ) & = & \mathrm{E}( X_i ) \label{glosa} \\ 
 \mathrm{V}( Y_i ) & = & \mathrm{V}( X_i ) + \mathrm{V}( {\Phi_\mathrm{m}}_i ) +
       \mathrm{V}( {\Phi_\mathrm{s}}_i ) \label{glosb}\\
 \mathrm{C}( Y_i , Y_j ) & = & \mathrm{C}( X_i , X_j ) + 
                      \mathrm{C}( {\Phi_\mathrm{m}}_i , {\Phi_\mathrm{m}}_j ) +
    \mathrm{C}( {\Phi_\mathrm{s}}_i , {\Phi_\mathrm{s}}_j ) \label{glosc}
\end{eqnarray}
Eliminating $\mathrm{C}( X_i , X_j )$ and $\mathrm{E}( X_i)$ 
from Eqs.~(\ref{globa}--\ref{globc}) by means of 
Eqs.~(\ref{glosa}--\ref{glosc}), we obtain a linear system of 
three equations in $\alpha$, $\beta$ and $\gamma$, 
whose solution is: 
\begin{eqnarray}
 \alpha & = & \Delta^{-1} \cdot 
   \left[ \mathrm{C}( Y_2 , Y_3 ) -
          \mathrm{C}( {\Phi_\mathrm{m}}_2 , {\Phi_\mathrm{m}}_3 )
   - \mathrm{C}( {\Phi_\mathrm{s}}_2 , {\Phi_\mathrm{s}}_3 ) \right]
      \cdot \nonumber \\
    &   &  \cdot \left[ \mathrm{C}( Y_1 , Y_2 )
   - \mathrm{C}( {\Phi_\mathrm{m}}_1 , {\Phi_\mathrm{m}}_2 )
   - \mathrm{C}( {\Phi_\mathrm{s}}_1 , {\Phi_\mathrm{s}}_2 ) \right] +     \nonumber \\ 
   & - & \Delta^{-1}  \cdot
   \left[ \mathrm{C}( Y_1 , Y_3 ) - \mathrm{C}( {\Phi_\mathrm{m}}_1 , {\Phi_\mathrm{m}}_3 ) -
  \mathrm{C}( {\Phi_\mathrm{s}}_1 , {\Phi_\mathrm{s}}_3 ) \right] \cdot \nonumber \\
   &   &             \cdot \left[ \mathrm{V}( Y_2 ) - \mathrm{V}( {\Phi_\mathrm{m}}_2 ) 
   - \mathrm{V}( {\Phi_\mathrm{s}}_2 ) \right] \label{fita} \\
 \beta  & = & \Delta^{-1}  \cdot
   \left[ \mathrm{C}( Y_1 , Y_2 ) - \mathrm{C}( {\Phi_\mathrm{m}}_1 , {\Phi_\mathrm{m}}_2 ) 
   - \mathrm{C}( {\Phi_\mathrm{s}}_1 , {\Phi_\mathrm{s}}_2 ) \right] \cdot \nonumber \\
   &   &  \cdot \left[ \mathrm{C}( Y_1 , Y_3 ) - \mathrm{C}( {\Phi_\mathrm{m}}_1 , {\Phi_\mathrm{m}}_3 ) 
   - \mathrm{C}( {\Phi_\mathrm{s}}_1 , {\Phi_\mathrm{s}}_3 ) \right] +      \nonumber \\
   & - &  \Delta^{-1} \cdot 
   \left[ \mathrm{C}( Y_2 , Y_3 ) - \mathrm{C}( {\Phi_\mathrm{m}}_2 , {\Phi_\mathrm{m}}_3 ) 
   - \mathrm{C}( {\Phi_\mathrm{s}}_2 , {\Phi_\mathrm{s}}_3 ) \right] \cdot \nonumber \\
   &   &             \cdot \left[ \mathrm{V}( Y_1 ) - \mathrm{V}( {\Phi_\mathrm{m}}_1 ) 
   - \mathrm{V}( {\Phi_\mathrm{s}}_1 ) \right] \label{fitb} \\
  \gamma & = & \mathrm{E}( Y_3 ) - \alpha  \mathrm{E}( Y_1 ) - 
     \beta \mathrm{E}( Y_2 ) \label{fitc} 
\end{eqnarray}
 where we put:
\begin{eqnarray}
 \Delta & = & \left[ \mathrm{C}( Y_1 , Y_2 ) - \mathrm{C}( {\Phi_\mathrm{m}}_1 , {\Phi_\mathrm{m}}_2 )   
   - \mathrm{C}( {\Phi_\mathrm{s}}_1 , {\Phi_\mathrm{s}}_2 ) \right]^2 + \nonumber \\
        & - & \left[ \mathrm{V}( Y_1 ) - \mathrm{V}( {\Phi_\mathrm{m}}_1 ) - \mathrm{V}( {\Phi_\mathrm{s}}_1 ) \right] \cdot \nonumber \\ 
    & \cdot & \left[ \mathrm{V}( Y_2 ) - \mathrm{V}( {\Phi_\mathrm{m}}_2 ) - 
    \mathrm{V}( {\Phi_\mathrm{s}}_2 ) \right] 
                \label{fitd}
\end{eqnarray}
On increasing the sample size, the expressions  obtained by
substituting in the previous equations the terms ${ \mathrm{C}( Y_i ,
Y_j ) }$ with their unbiased estimates ${ \hat{ \mathrm{C} }( Y_i ,
Y_j ) }$ (hereafter $\hat{\alpha}$, $\hat{\beta}$  and
$\hat{\gamma}$), will approximate more and more closely the `true
coefficients' $\alpha$, $\beta$ and $\gamma$. The quantities
$\hat{\alpha}$, $\hat{\beta}$ and $\hat{\gamma}$ will thus furnish
unbiased, asymptotically normal, estimates of $\alpha$, $\beta$ and
$\gamma$. In App. B the analytical formulae of the relative CM
components are also derived. \\ If \yi~ are normally distributed, the
unbiased estimators for the moments of \yi~ are defined by the
following usual expressions:
\begin{eqnarray}
  { \hat{ \mathrm{E} }( Y_j )} & = & \frac{1}{N} \cdot
   \sum_{i=1,N} y_{j_i} \label{nesta} \\
  { \hat{ \mathrm{V} }( Y_j )} & = & \frac{1} {N-1} \cdot
   \sum_{i=1,N} \left( y_{j_i} - \overline{ Y_j} \right)^2 \label{nestb} \\
   { \hat{ \mathrm{C} }( Y_j , Y_k ) } & = & 
   \frac{1}{N-1} \cdot \sum_{i=1,N}
   \left( y_{j_i} - \overline{ Y_j} \right) \cdot
   \left( y_{k_i} - \overline{ Y_k} \right) \label{nestc}
\end{eqnarray}
The fitting procedure based on Eqs.~(\ref{fita}--\ref{fitd})  and  
(\ref{nesta}--\ref{nestc}) will be indicated hereafter by the acronyms
$MIST$, 
{\it Measurement errors and Intrinsic Scatter Three dimensional} fit. \\

\subsection{Deriving the coefficients of the FP.}

To discuss the application of Eqs.~(\ref{fita}--\ref{nestc}) to the
derivation of the FP, it is important to remark the following points:
\begin{description}
  \item[\op.]  The terms $\mathrm{V}( {\Phi_\mathrm{s}}_3 )$ 
    and  $\mathrm{V}( {\Phi_\mathrm{m}}_3 )$ do not appear in
    Eqs.~(\ref{fita}--\ref{fitd}). The 
    estimators of $\alpha$, $\beta$ and  $\gamma$ are therefore 
    independent of the scatter along the dependent variable.
  \item[\os.] Setting equal to zero the other components of the CMs of
    \fmi~ and \fsi~ in Eqs.~(\ref{fita}--\ref{fitd}) and using     
    Eqs.(\ref{nesta}--\ref{nestc}), we obtain exactly the expressions 
    of the $\mathrm{OLS}_{ Y_3} $ estimators, where the subscript      
    $Y_3$ is used to indicate the dependent variable in the 
    fit\footnote{The OLS method can be thus regarded, in this respect, 
    as a particular case of the MIST fit, in which the scatter is 
    completely ascribed to the direction of the dependent variable.}.      
    If, on the other hand, every variable is affected by dispersion, the
    quantities $\mathrm{C}( Y_1 ,Y_2 )$ would give a biased estimate
    of $\mathrm{C}( X_1 , X_2 )$ as shown by 
    Eqs.~(\ref{globb}, \ref{globc}).      To correct for the bias
    the CM components of \fmi~ and      \fsi~ are needed (see AkB96).
  \item[\ot.] As follows from \op~ and \os, to recover the mean relation
    by Eqs.~(\ref{fita}--\ref{fitd}) we have altogether ten degrees of
    freedom, given by five elements of the CMs of  \fmi~
    and \fsi~ respectively.
  \item[\oq.] Since Eqs.~\ref{relsca} have the same dependency on the      
    three variables, \emph{ the best-fit coefficients obtained by means 
    of Eqs.~(\ref{fita}--\ref{fitd}) are independent of the choice of 
    the dependent variable}.
  \item[\oqu.] New unbiased estimators of the coefficients in
Eq.~(\ref{relmed}) can be defined by taking some average of the slopes of
the three planes determined  by Eqs.~(\ref{fita}--\ref{fitd}) using each of
the \yi~ as dependent variable. We can define a {\it
bisector plane} (see also Graham \& Colless \cite{grc96}), whose slopes are
given by the vectorial sum of the normal vectors to the three planes
obtained by Eqs.~(\ref{fita}--\ref{fitd}), and whose zero point is obtained
by Eq.~(\ref{fitc}). The advantage of  the `bisector' fit (hereafter \erbi~
fit) consists in the larger effectiveness (i.e. smaller variances for given
sample size) of the relative estimators (see Sect.~3.2, and IFA90 for the
two-dimensional case).
  \item[\osx.] There are often large systematic in-homogeneities between 
     different samples of data, due to differences in the procedures 
     adopted to derive them (see e.g. Smith et al. \cite{slh97}). 
     To account for such in-homogeneities, the least squares fits 
     (ORLS and OLS) are often performed using robust estimators. The 
     robust statistics can also be implemented to the MIST regression, 
     by using robust estimators for the moments of \yi~ in 
     Eqs.~(\ref{fita}--\ref{fitd}).
\end{description} 

From the  derivation of the equations in the previous Section, it
follows that, in order to recover the coefficients of the mean relation,
$\alpha$, $\beta$ and $\gamma$, some requests must be satisfied:
\begin{description}
   \item[\hyp:] \xxi, \fmi~ and \fsi~ are not mutually correlated;
   \item[\hys:] estimates of the CM components of \fmi~ are known;
   \item[\hyt:] estimates of the CM components of \fsi~ are available;
   \item[\hyq:] unbiased estimates of the CM components of \yi~ are known.
\end{description}

The hypothesis \hyp~ is practically equivalent to assume that the
measurement errors and the intrinsic scatter on the FP variables do not
depend on the `location' on the FP. This is not the case for the
measurement uncertainties on the observed parameters 
and, {\it a priori}, it could be not true even for the
intrinsic dispersion. However Eqs.~(\ref{glosb}--\ref{fitd}) 
continue to
be valid by simply substituting the CM components of \fmi~ and \fsi~
with their expected values as a function of the `position' on the plane
(see App. A and B for details).

Concerning the measurement uncertainties (hypothesis \hys),
Eqs.~(\ref{fita}--\ref{fitd}) can still be used by adopting averaged values of
$\mathrm{C}( {\Phi_\mathrm{m}}_i , {\Phi_\mathrm{m}}_j )$. To this aim, we
suggest to proceed as follows. The quantities $\mathrm{V}( {\Phi_
\mathrm{m}}_i )$ can be estimated as the square of the mean
error\footnote{Notice that the mean errors are often the only significant
available estimates.} on the parameters $\mathrm{ {Y_m}_i }$. For what
concerns $\mathrm{C}( {\Phi_\mathrm{m}}_i, {\Phi_\mathrm{m}}_j )$, the only
term that does not vanish, is $\mathrm{C}( \log r_\mathrm{e}, \mu_\mathrm{e})$
due to the correlation between the uncertainties in  $\log r_\mathrm{e}$ and
$\mu_\mathrm{e}$. The quantities $\mathrm{C}( {\Phi_\mathrm{m}}_i,
{\Phi_\mathrm{m}}_j )$ are never given in literature, so that one is forced to
make some approximations. Since $\Delta \log r_\mathrm{e} \approx \alpha \!
\cdot \! \Delta \mu_\mathrm{e}$ with $\alpha \approx 0.3$ (see JFK96), we can
set $\mathrm{C}( \log r_\mathrm{e}, \mu_\mathrm{e}) \approx \alpha \! \cdot \!
\mathrm{V}( \mu_\mathrm{e} )$. The uncertainties introduced by this
approximation will be discussed in Sect.~4.

Assumptions \hyt~ and \hyq, cannot be satisfied because we do not have a
physical model of the probability distribution of E galaxies in the
space of the observed quantities. 
It is thus necessary to introduce some simplifying assumptions:
\begin{description}
   \item[\hytp]: on the basis of \op~ we can only
      assign all the intrinsic dispersion to the dependent
      variable.
   \item[\hyqp]: the only possible simple assumption is that
      \yi~ are normally distributed (see Eqs.~\ref{nesta}--\ref{nestc}).
\end{description}
Both these assumptions introduce a `bias' in the estimate of $\alpha$,
$\beta$ and $\gamma$, so that \emph{ the coefficients $a$, $b$ and
$c$ in Eq.~(\ref{fpeq}) do not necessarily coincide with $\alpha$,
$\beta$ and $\gamma$ }. Moreover, because of \hytp, the estimates obtained 
by Eqs.~(\ref{fita}--\ref{fitd}) with a different choice of the dependent 
variable do not correspond to the same statistical model and so do not 
define the same plane. 

In order to better understand the bias introduced by \hytp~ and
\hyqp~, we used numerical simulations of the FP. The simulations were
constructed giving a scatter around the plane in the \ls~ direction by
a normal RV (see Sect.~3.1 for details). The variance of $\log
\sigma_0$ has been varied in the range of values obtained for the
$\log \sigma_0$ scatter of different samples of galaxies (see
Sect.~4.1). The FP coefficients have been derived for each simulation
by the MIST fits, with assumptions \hytp~ and \hyqp, using each of
\yi~ as the dependent variable (hereafter $\mathrm{MIST}_{Y_i}$ fits),
and by the MIST$_{\mathrm{BIS}}$. We used the estimates of
$\mathrm{\mathrm{C}( {\Phi_\mathrm{m}}_i , {\Phi_\mathrm{m}}_j )}$
obtained by typical values of the uncertainties on FP parameters (0.03
in \lr, 0.1 in \br~ and 0.03 in \ls).

\begin{figure}
  \resizebox{\hsize}{7cm}{\includegraphics{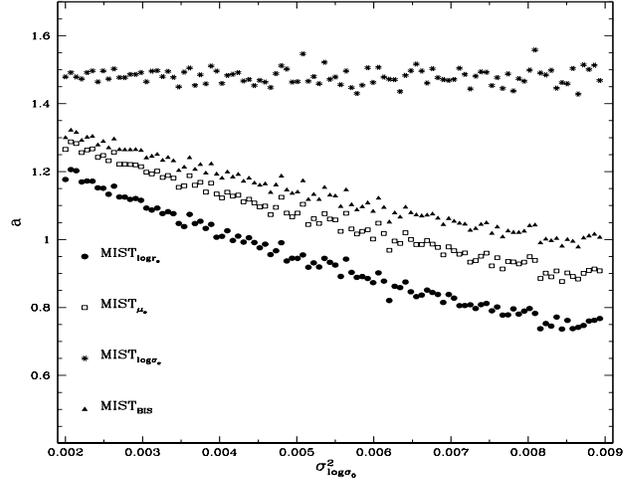}}
  \caption[]{ Coefficient $a$ of FP simulations against the $\log \sigma_0$
   variance, $\sigma^2_{\log \sigma_0}$. The size of the
   simulations is $N = 3000$. The symbols correspond
   to different fitting methods as shown in the lower left.
}
  \label{disp}
\end{figure}

In Fig.~\ref{disp}, we plot the coefficient $a$ against $\sigma^2_{\log
\sigma_0}$ (similar results are obtained for $b$ and $c$). The values of the
FP coefficients obtained by the various fitting methods turn out to be
systematically different, by an amount that increases with the scatter. 
An estimate of this difference for the MIST fits will be given in
Sect.~4 by comparing the FPs of different clusters of galaxies.

\begin{table*}
\caption[]{ Values of $a$, $b$ and $c$ of a FP simulation with 
intrinsic dispersion 'known' (see text). The sample size is  $N = 3000$.
Column 1: FP coefficients $a$, $b$ and $c$. Columns 2, 3, 4 and 5: values 
obtained by the different MIST fits (seesubscripts in the top row) with 
corresponding uncertainties (1$\sigma$ intervals).  
 \vspace{0.25cm}}
\hspace{0.1cm}
\centering
\begin{tabular}{|c|c|c|c|c|}
\hline
 & & & & \\

 & $\mathrm{log r_e}$  & $\mathrm{ \mu_e }$ & $\mathrm{ log \sigma_0 }$ & 
   BIS   \\
 & & & & \\
\hline
 & & & & \\
 $a$ &  $1.26 \pm 0.02$  & $1.28 \pm 0.02$   & $1.28 \pm 0.01$ & 
   $1.28 \pm 0.01$ \\
 $b$ & $0.310 \pm 0.002$ & $0.308 \pm 0.002$ & $0.309 \pm 0.002$ &
   $0.309 \pm 0.002$\\
 $c$ & $-8.27 \pm 0.07$  & $-8.23 \pm 0.07$  & $-8.20 \pm 0.07$ & 
   $-8.24 \pm 0.06$\\
 & & & & \\
\hline
\end{tabular}
\label{diffit}
\end{table*}

The assumption \hyqp~ introduces likely the same amount of bias in the
various fits. To illustrate this point, let us consider an example in which
the CM components of the intrinsic dispersion are 'known'. We will assume
$\mathrm{ \mathrm{V}( {\Phi_\mathrm{m}}_{\log \sigma_0} ) + \mathrm{V}(
{\Phi_\mathrm{s}}_{\log \sigma_0} ) } =  \mathrm{ \mathrm{V}(
{\Phi_\mathrm{m}}_{\log r_e} ) + \mathrm{V}( {\Phi_\mathrm{s}}_{\log r_e }
) } =   0.0035$ (a scatter of $\sim 14 \%$ in $\mathrm{r_e}$ and
$\sigma_0$)\footnote{This amounts to a 15\% scatter in the $\sigma_0$
direction, i.e. to the mean value of the FP scatter of clusters, see
Sect.~4.} and that the other CM terms of \fmi~ and \fsi~ vanish. We now
create a FP simulation, and derive $a$, $b$ and $c$ by the MIST fits, using
the assigned values of the CM components of \fsi~ and \fmi, and making
therefore \emph{ only} the hypothesis \hyqp.
The results of the fits are shown in Tab.~\ref{diffit}: the FP
coefficients are practically independent of the fitting method.

We then conclude that the differences of Fig.~\ref{disp} are due only to the
assumption \hytp: \emph{ the lack of a model for the intrinsic dispersion
of the FP variables is thus at the origin of the dependency of the FP
coefficients on the fitting procedure}. 
The larger is the scatter around the plane, the larger will be the bias
due to the fitting method.

\section{Uncertainties on the FP coefficients.} 

Two kinds of methods are usually adopted to estimate the
uncertainties on fit coefficients: theoretical methods and re-sampling
techniques. 

The `theoretical uncertainties' are obtained from the analytical
expression of the variances of the estimators (e.g. IFA90 and Feigelson \&
Babu \cite{feb92}). By their nature, these estimates are valid only 
asymptotically, i.e. for large sample sizes. 

When analytical formulae are not available, or when the sample is 
small, re-sampling procedures are adopted. 
The statistics of interest is calculated for various `pseudo-samples' 
drawn from the original data set. The uncertainties are then derived
from the distribution of pseudo-values. The main re-sampling procedures 
are known as `jackknife' (see Quenouille \cite{que49} and 
Tukey \cite{tuk58}) and `bootstrap' (see Efron \cite{efr79} and 
Efron \& Tibshirani \cite{eft86}).
In the jackknife, one point is extracted in sequence from the original 
data set, so that a number of pseudo-samples equal to the sample size is 
constructed.
In the bootstrap, random samples are drawn by replacement from the actual 
data set.

\begin{figure}
  \resizebox{\hsize}{7cm}{\includegraphics{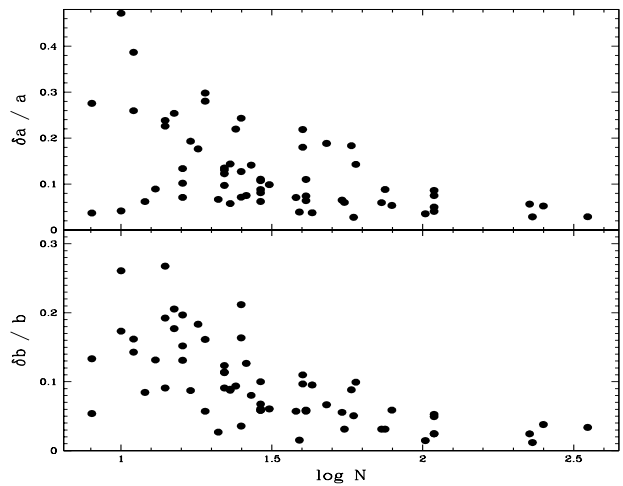}}
  \caption[]{
   Relative uncertainties $\delta a / a$ and $\delta b / b$ on FP
   slopes against the sample size (see text). The values are 
   drawn from literature.   
   Sources of the uncertainties and relative sample sizes 
   (within square brackets): 
   Busarello et al. (\cite{bcc97}) [40]; 
   de Carvalho \& Djorgovski (\cite{dcd92}) [31 40 55 58]; 
   D'Onofrio et al. (\cite{dcz97}) [25]; 
   Dressler et al. (\cite{dlb87}) [43]; 
   Graham (\cite{gra98}) [7 18 25 25];
   Graham \& Colless (\cite{grc96}) [26]; 
   Hudson et al. (\cite{hls97}) [352]; 
   J\o rgensen et al. (\cite{jfk96}) [79 22 24 14 10 14 29 8 19 226 109 109 109
   109 41 41]; 
   Mobasher et al. (\cite{mga99}) [48]; 
   Pahre et al. (\cite{pdc95}) [8 10 12 13 16 59]; 
   Pahre et al. (\cite{pcd98}) [251 60 16 23 15 14 15 17 11 19 11 27]; 
   Prugniel \& Simien (\cite{prs94}) [102];
   Prugniel \& Simien (\cite{prs96}) [231 39 21];
   Recillas-Cruz et al. (\cite{rcs90}) [29 29 29];
   Recillas-Cruz et al. (\cite{rcs91}) [22 22 22];
   Scodeggio et al. (\cite{sgb98}) [38 41 54 75 73 29]. 
  }
  \label{ercoflet}
\end{figure}

Figure \ref{ercoflet} shows the relative uncertainties on the FP
coefficients $a$ and $b$, as drawn from literature, as a function of the
sample size $N$. 
{\it The large scatter of the values of the uncertainties is indicative of
the inconsistency between the methods used in the different works to
estimate the errors}.

In the following we will analyze the performance of the methods to estimate
the FP uncertainties with special regard to the r\^ole of the sample size.
The analysis will be performed by numerical simulations, which are
described in the next section.

\begin{figure*}
 \resizebox{\hsize}{13cm}{\includegraphics{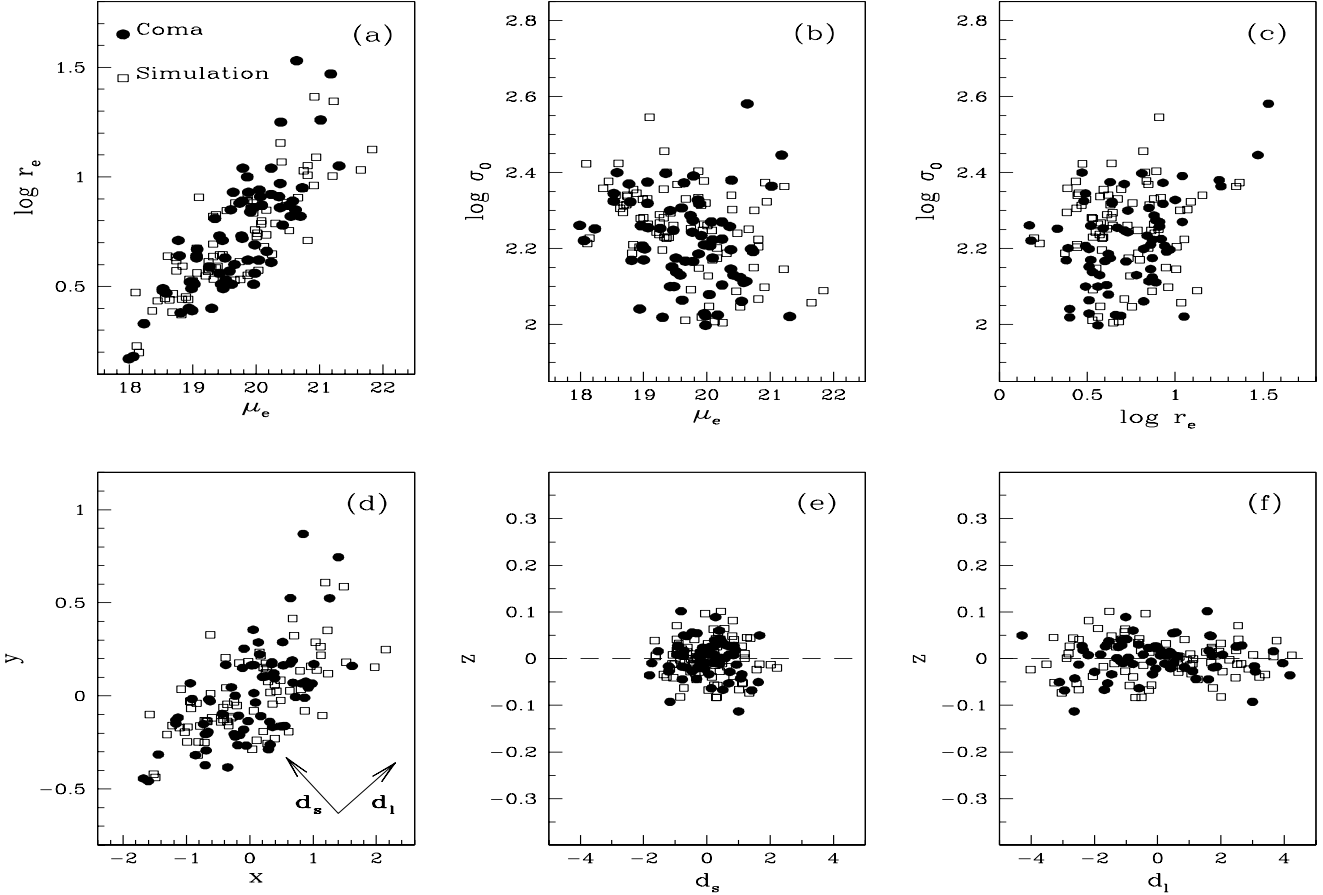}}
 \caption[]{Comparison of one of the FP simulations with the template
  sample. (a) Simulation and the Coma 
  photometric sample in the plane $\mathrm{(\mu_e , \log r_e)}$. 
  (b) Distribution in the plane
  $\mathrm{ (\mu_e , \log \sigma_0)}$. (c) Distribution in the plane
  $\mathrm{ (\log r_e , \log \sigma_0) }$. (d) Face-on view 
  of the FP. (e), (f) Long and short edge-on views of the FP. 
  The plotted simulation has the same size as the Coma sample.}
\label{simtem}
\end{figure*}
\subsection{The simulation algorithm.}

The simulations consist in distributions of points extracted from 
a common `parent distribution'.
To derive the parent population we based on the distribution in the
parametric space of the galaxies in the Coma cluster (hereafter the 
`template' sample).
This choice is mandatory since Coma is the only cluster
with FP parameters available for a large number of galaxies.

The simulation algorithm consists in the following.
\begin{description}
 \item At first, we derived a parent distribution of E galaxies
      in the plane (\br,\lr), using the photometric data in the 
      Gunn $\mathrm{r}$ band  by JFK95a for 146 
      galaxies in Coma. 
      The sample is complete out to Gunn $\mathrm{r} \simeq 15.5$ mag.
      The distribution with respect to \br~ was described by 
      a normal RV.
      The \br~ interval of the template was binned,  
      and the mean value (MV) and standard deviation (SD)
      of the \lr~ distribution derived for each bin.
      The MVs and the SDs were then fitted  with respect to the 
      central values of \br~ by polynomials of suitable order, and 
      the best-fit curves were used to interpolate a value (and a 
      scatter) of \lr~ to each value of \br.
 \item Using $\sigma_0$  by J\o rgensen et al. (\cite{jfk95b},
      hereafter JFK95b) for 75 
      galaxies of the photometric sample (see Sect.~4), we determined 
      the FP coefficients and the root mean square (\smr) of \ls~ residuals,
      $s_{log \sigma_0}$, by the \fls~ fit. 
      We found $a \simeq 1.50$, $b \simeq 0.328$, $c \simeq -9.1$ and 
      $s_{log \sigma_0} \simeq 0.048$. These quantities 
      were used to assign to the points a value and a scatter 
      with respect to the variable \ls. 
\end{description} 

One of the simulated samples is compared to the template in Fig. 
\ref{simtem}, to show how the simulations resemble very well 
the distribution of Coma galaxies. 

\subsection{Estimating the uncertainties.} 
We derived the `true uncertainties' on the FP coefficients as a function of
the sample size in the following way.

\begin{figure}
 \resizebox{\hsize}{7cm}{\includegraphics{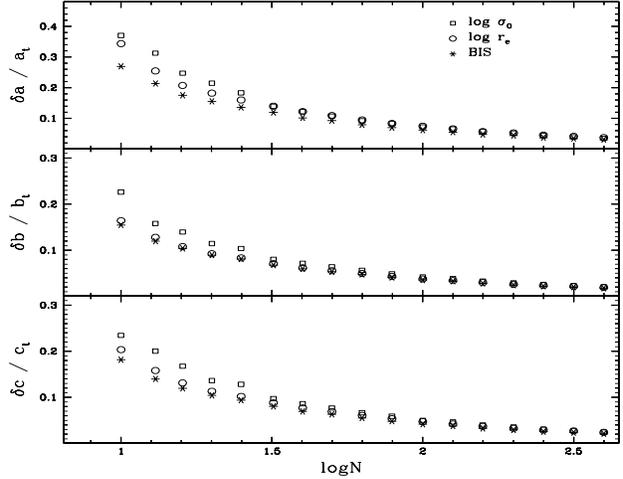}}
 \caption[]{ `True relative uncertainties' on the FP coefficients 
  against the logarithm of the sample size $N$. 
  The various MIST fits are plotted by different symbols as
  shown in the upper right-hand corner. The \erlr~ and \erbr~ fits
  are completely consistent, so that for clarity only the 
  first one is plotted.
  The axes scales are the same as in Fig.~\ref{ercoflet}.    
}
 \label{ercofver}
\end{figure}

FP simulations of fixed size $N$ were constructed and the
coefficients $a$, $b$ and $c$ determined for each sample. The 'true
uncertainties', $\delta a$, $\delta b$ and $\delta c$ were determined
from the distributions of $a$, $b$ and $c$ by using 1$\sigma$ standard
intervals. To obtain estimates independent of the 'fitting scale', we
divided $\delta a$, $\delta b$ and $\delta c$ by the 'true' coefficients 
of the FP, $a_t$, $b_t$ and $c_t$, derived from a simulation of large size. 
In Fig.\ref{ercofver} we plot $\delta a / a_t$, $\delta b / b_t$ and 
$\delta c / c_t$ against the logarithm of the sample size $N$ 
for the various MIST fits. To allow a direct comparison with 
Fig.~\ref{ercoflet} the same range of $N$ has been plotted.

It is apparent that the true intervals depend on the fitting method and
that the most effective fit (i.e. lower values of $\delta a / a$ and
$\delta b / b$ for fixed $N$) is the \erbi~ method, in agreement
with what found by IFA90 for the bisector line. The
largest variances are obtained for the \erls~ fit. It is also worth to
notice that, by changing the scatter along the $\sigma_0$ direction in
the simulations, the curves shown in the figure undergo just a
translation in the $y$ direction, without any change in shape.
In the case shown in Fig.~\ref{ercofver} we adopted the MV 
of the \smr~ of \ls~ residuals of various clusters of galaxies 
(see Sect.~4). 

Figure \ref{ercofver} can be used as a ready tool to state the 
number of galaxies necessary to achieve a given accuracy on the FP.

Concerning the zero point of the FP, it is important to remark that
the uncertainties plotted in Fig.\ref{ercofver} do not represent the 
estimates of usual interest. For all the applications of the FP
zero point (i.e. distance determinations, constraining of cosmological
and evolution parameters), the uncertainties on $c$ are derived with the
hypothesis that the FP slopes are exactly determined. The errors on $c$
are then given as $1 / \sqrt{N}$ multiplied by the \smr~ of the 
dependent variable residuals. Such estimates are generally smaller 
than $\delta c / c_t$.

For this reason, in the following we will focus our analysis on the
uncertainties of the FP slopes.

Although the comparison of Figs.~\ref{ercoflet} and \ref{ercofver} does 
not show an evident disagreement, for $\log N \widetilde{<} 1.8$ 
($N \widetilde{<} 60$) the values reported in literature appear 
almost as a scatter diagram.
Starting from this remark, we now analyze the performances 
of the different methods used to estimate the uncertainties. 

\subsubsection{Theoretical methods.} 

Although statistics allows to prove the asymptotic validity of 
variance estimators, it does not furnish an estimate of the 
`minimum sample size' for the theoretical formulae to be valid.
Since such an estimate will generally depend on the 'shape' of the 
parent population, it should be obtained each time by using
simulation methods. 

\begin{figure}
 \resizebox{\hsize}{7cm}{\includegraphics{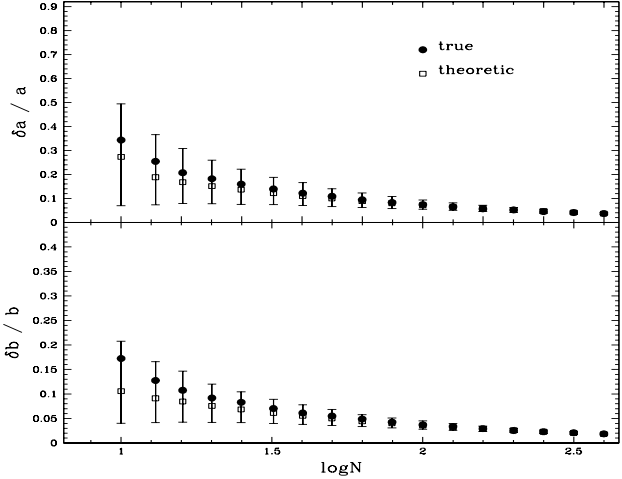}}
 \caption[]{ Comparison of theoretic and true relative uncertainties 
  on the FP slopes (see text). The different symbols are explained in the 
  upper right-hand corner. We adopted the \erlr~ fit. 
}
 \label{ertheor}
\end{figure}

To test the performance of theoretical variance estimators for the FP
coefficients we apply the results of Sect.~2. FP simulations of fixed
size $N$ were constructed and theoretical relative uncertainties
(1$\sigma$ standard intervals) on $a$ and $b$ derived for each sample
by Eqs.~(B17, \ref{fita} -- \ref{nestc}). In Fig.~\ref{ertheor}, the
MVs of the theoretical estimates, $\delta a / a$ and $\delta b /b$,
with corresponding `error bars', are plotted against the logarithm of
$N$. The error bars were obtained by connecting the 5th and 95th
percentiles\footnote{ The $p$th percentile of a given distribution 
is the point that exceeds (100-p)
ascending order (see e.g. Beers et al. \cite{bfg90}). \label{notperc} } of the
$\delta a / a$ and $\delta b /b$ distributions. For comparison, we
also plot the 'true uncertainties' on the slopes as derived from the
simulations. As expected, the larger the sample size, the better it is
the agreement between theoretical and actual values. For sample sizes
smaller than $N_{\mathrm{min}} \approx 30$ ($\log N_{\mathrm{min}} =
1.5$), we see however that theoretical formulae (I) largely scatter
(up to $40 \%$ for $a$ and $15 \%$ for $b$)  and (II) increasingly
underestimate the 'true' values (see also IFA90 and Feigelson \& Babu
\cite{feb92}). 

To discuss in more detail the point II, we calculated for each sample
the coefficient a, the theoretical uncertainty ${\delta a}$, and the
`discrepancy' $d_a = |a - a_\mathrm{t}| / {\delta a}$, where the
actual value, $a_\mathrm{t}$, of the coefficient $a$ was derived by
fitting a simulation of large size. We then determined the fraction of
simulations, $f_{\Delta} $, with $d_a$ greater than a fixed value
$\Delta$. The calculation was iterated by varying the sample size.

\begin{figure}
  \resizebox{\hsize}{7cm}{\includegraphics{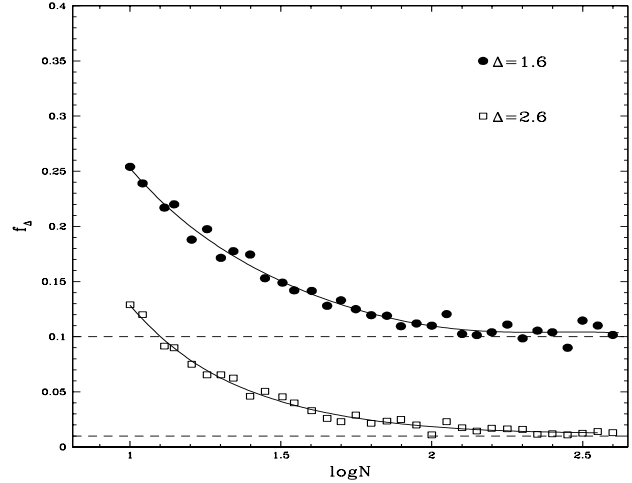}}
  \caption[]{ Fraction of non-consistent simulations, 
    ${f_\Delta}$, against the sample size $N$ (see
    text). Circles and rectangles correspond to different values of
    $\Delta$ as shown in the upper right-hand corner.
    The solid lines were obtained by interpolation. 
    The dashed lines show the asymptotic values of the curves.
}
  \label{inc}
\end{figure}

In Fig.~\ref{inc}, we plot ${f_\Delta}$ against $\log N$ for two
different values of $\Delta$. We chose $\Delta = 1.6$ that, for a
normal distribution, defines a confidence level (CL) of $10 \%$, and
$\Delta = 2.6$, corresponding to a $1 \%$ CL. If Eqs.~(\ref{nesta} --
\ref{nestc}) worked well, on the average, for every value of $N$, the
fraction of not-consistent samples would be independent of $N$ and
determined by the CL corresponding to the value of $\Delta$.
Figure~\ref{inc} shows that for $ N > N_{\mathrm{min}} \approx 30$
($\log N_{\mathrm{min}} \approx 1.5 $) ${f_\Delta}$ is in good
agreement (within $\sim$5\%) with the expected CLs. For sample sizes
smaller than $N_{\mathrm{min}}$, the theoretical formulae do not
furnish reliable estimates of the desired CLs: the curves of
Fig.~\ref{inc} increase steeply. The same result was obtained for the
coefficient $b$, and by varying the MIST fit. We also found that the
estimate of $N_{\mathrm{min}}$ is largely independent of the
simulation parameters: the same estimate was obtained using the FP
coefficients and the $\log \sigma_0$ dispersion of the samples
studied in Sect.~4.

Figure \ref{inc} suggests that in order to obtain a given confidence
level, the uncertainties on the FP coefficients of small samples
should be estimated using a $\sigma$ interval dependent on $N$. For
instance, to obtain a $10 \%$ CL with $N = 12$ ($\log N = 1.1$), an
effective interval of 2.6$\sigma$ should be used. However, it turns
out that the smaller the sample size with respect to 
$N_{\mathrm{min}}$, the stronger is the dependence of $f_\Delta$ on
the adopted fitting method and, what is more critical, on the
simulation parameters. For instance, the fraction of not-consistent
samples with $N = 10$ varies by $\sim 6 \%$ by varying
the $\log \sigma_0$ scatter
in the simulations.
   
We conclude that for $N \widetilde{<} 30$ the theoretical formulae are
not reliable. Although the desired confidence intervals can be roughly
obtained by using effective, suitably tested, standard intervals, the
individual estimates can be significantly different, up to $\sim 20
\%$ for $a$ and $\sim 15 \%$ for $b$, with respect to the true values.

\subsubsection{Re-sampling procedures.}

The hypothesis underlying the use of re-sampling methods is that the
available data set furnishes a good approximation to the parent
population. The statistics of interest is calculated for various
pseudo-samples drawn from the actual data set. If this `sampling
hypothesis' holds, the distribution of pseudo-values coincides with
the `true' one and the confidence intervals can be accurately estimated at
the cost of some computing time (see e.g. Efron \& Tibshirani
\cite{eft86}). However, the smaller the sample size, the larger is the
probability that the actual sample gives a poor representation of the
parent population. In order to derive a `minimum sample size' for the
re-sampling methods to be reliable, numerical simulations have to be
employed.

On the basis of the analysis of the previous Section, we studied the
performance of re-sampling uncertainties on the FP coefficients by
testing the use of the bootstrap method. The results that follow were
found to be largely independent of the simulation parameters, of the
actual FP coefficients, and of the fitting method.

\begin{figure} 
\resizebox{\hsize}{7cm}{\includegraphics{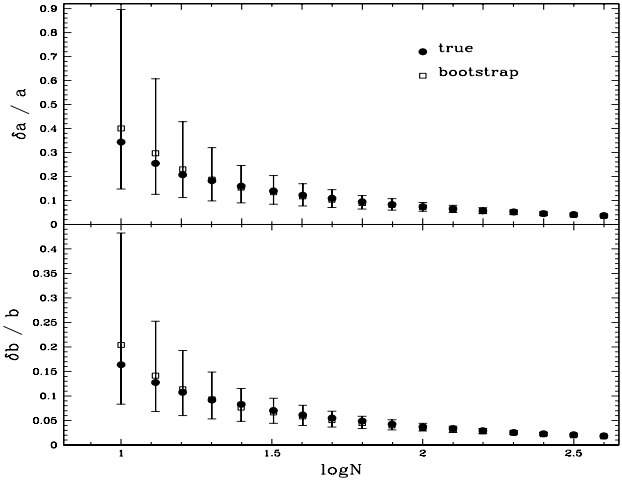}}  \caption[]{
Comparison of `bootstrap' and 'true' relative uncertainties   on the
FP slopes (see text). Different symbols have been used   as indicated
in the upper right-hand corner. We adopted the   \erlr~ fit. The axes
scales are the same of Fig.~\ref{ertheor}. }  \label{erboot}
\end{figure}

FP simulations of fixed size $N$ were constructed. For each sample, we
determined the FP slopes and the bootstrap uncertainties ${\delta a}$
and ${\delta b}$ by using 1$\sigma$ standard intervals with
$N_\mathrm{r}=2000$ pseudo-samples (by further increasing
$N_\mathrm{r}$ the following results do not change). In
Fig.~\ref{erboot}, the MVs of the bootstrap relative uncertainties,
$\delta a / a$ and $\delta b / b$, with corresponding error bars (see
Sect.~3.2.1), are plotted against the logarithm of the sample size and
compared to the true values. The error bars are given by $10 \%$
percentile intervals\footnote{ A $p \%$ percentile interval is given
by the difference between the $p/2$th and the $(100 - p/2)$th
percentiles (see note \ref{notperc}) of the distribution of interest.
}.

\begin{figure}
  \resizebox{\hsize}{7cm}{\includegraphics{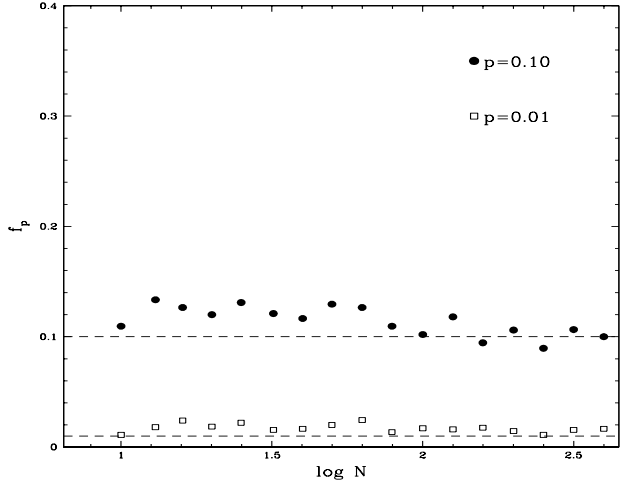}}
  \caption[]{ Fraction of non-consistent simulations, $\mathrm{ {f_p} }$, 
    against $\log N$ (see text). The plot is referred to the 
    coefficient $b$. Circles and rectangles correspond to $100p \%$ 
    percentile intervals with $\mathrm{p} = 0.10$ and $\mathrm{p}=0.01$ 
    (see upper right-hand corner). The dashed lines show the desired 
    CLs. The axes scales are the same as in Fig.~\ref{inc}.
}
  \label{ncboot}
\end{figure}

As shown in the figure, the bootstrap method gives a good measure of the
average uncertainties, but a very poor, largely scattered, representation of
the actual errors. \\
Only for $N \widetilde{<} 15$ ($\log N \widetilde{<} 1.1 - 1.2$) the MVs of
the re-sampling standard errors appear to differ from the true uncertainties.
As a matter of fact, it turns out that this difference is due to the use of
1$\sigma$ standard intervals to estimate the bootstrap confidence intervals.
For small samples, in fact, the distribution of pseudo-values is significantly
different from a normal one, so that the desired CLs must be obtained by
non-parametric estimates\footnote{  I.e. by confidence intervals independent of
any assumption on the  shape of the probability density of the statistics of
interest. } (see Efron \& Tibshirani \cite{eft86} and Efron \cite{efr87}). To
illustrate this point, we tested the use of bootstrap percentile intervals
proceeding as in the previous Section (Fig.~\ref{inc}). For each sample size,
we derived the fraction of simulations, $f_\mathrm{p}$, that are not consistent
with the 'true' FP slopes. The calculation was iterated by adopting different
percentile intervals of the pseudo-values. To have a direct comparison with
Fig.~\ref{inc} we chose $100 \cdot p \%$ intervals with $p = 0.10$ and $p =
0.01$ respectively. In Fig.~\ref{ncboot}, the fraction of non-consistent
samples is plotted against the logarithm of the sample size. It turns out
that, on the average, the bootstrap allows to give very accurate estimates of
the true intervals. For every sample size, differences of only some percents
are found with respect to the desired CLs.

On the other hand, by looking at Fig.~\ref{erboot}, we notice that for
small samples the bootstrap uncertainties have a very large dispersion
with respect to true values. The scatter varies from $\sim 5 \%$ for
$N = N_{\mathrm{min}} \approx 30$ up to $60 \%$ for $N \approx
10$. Below $N \approx N_{\mathrm{min}}$ the FP parent population is
poorly represented, so that the single bootstrap estimates can be very
unsatisfactory.

\begin{figure}
  \resizebox{\hsize}{7cm}{\includegraphics{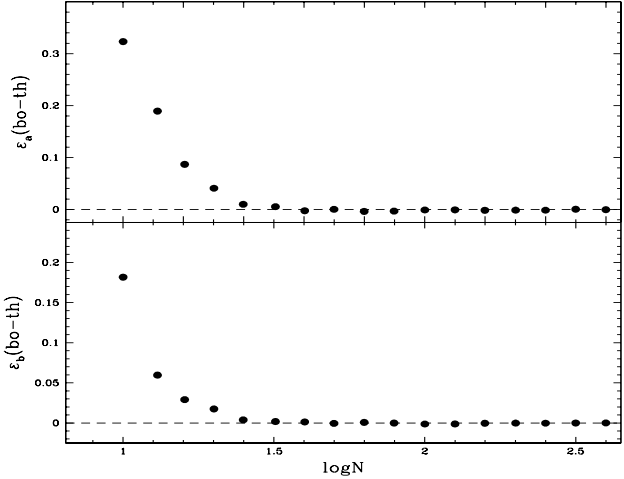}}
  \caption[]{ Differences, $\varepsilon_a(bo-th)$ and 
   $\varepsilon_b(bo-th)$,  of the 'scatters' of 
   bootstrap and theoretic uncertainties on $a$ and $b$, against 
   $\log N$.
}
  \label{boottheor}
\end{figure}

To have a comparison with the theoretical methods, we compared the
width of the error bars shown in Figs.~\ref{ertheor} and \ref{erboot}.
In Fig.~\ref{boottheor} we plot the difference of the error bars
against the logarithm of the sample size.

While for large samples theoretic and bootstrap uncertainties have a 
similar scatter, for $N \widetilde{<} N_{\mathrm{min}} \approx 30$ ($\log
N_{\mathrm{min}} \widetilde{<} 1.5$) the bootstrap errors become
increasingly less accurate. For $N \approx 10$, the scatter increases
up to $\sim 30 \%$ for $a$ and to $\sim 20 \%$ for $b$.
 
The conclusions are thus the following.

For large samples, both theoretic and bootstrap methods give accurate 
estimates of the uncertainties.

For $N \widetilde{<} 30$, both methods give estimates that can differ
significantly from the true values. The bootstrap is accurate on
the average but the uncertainties have a very large scatter. The
theoretical methods give values that are more precise, but 
systematically underestimated. 

\section{An application: the FP of cluster galaxies.}

The use of the FP as distance indicator is based on the hypothesis
that its slopes and thickness are `universal', i.e., independent of
the sample of galaxies.

The problem of the universality has been studied by JFK96 by deriving the 
FPs of ten clusters of galaxies.
Although the authors find that the FP coefficients are no
significantly different from cluster to cluster, they admit that the
samples are too small to get an accurate comparison.

Up to now,  the FP has been derived with a significant number of galaxies
only for Coma cluster ($N = 81$). Most of the other cluster samples have
$N \widetilde{<} 20 - 30$ (e.g. Dressler et al. \cite{dlb87}, JFK96,
Hudson et al. \cite{hls97}, Pahre et al. \cite{pcd98}).

On the basis of the discussion in Sects.~2 and 3, we try now to address the
question of the universality by comparing the FPs of different
clusters.

We chose the samples so that a) their size was as large as possible;
b) the FP parameters of the different samples were homogeneous; c) each
sample had photometric parameters in the same waveband and d) the
cluster were at $z\sim$ 0. The four prescriptions are needed
to a), b) make significant comparisons, c) avoid waveband
dependencies, and d) avoid galaxy evolution effects. The ten samples
surviving the above criteria are listed in Tab.~\ref{cluster}.

The FP parameters, \ls, \lr~ and \br~ were drawn from literature. The
photometry is in Gunn $\mathrm{r}$ for JFK data and in Kron-Cousins R band
for the other samples. For the present study, the differences of the two
wavebands can be completely neglected (see Pahre et al. 1998 and 
Scodeggio et al. 1998). Only galaxies with $\sigma_0 \ge 95 \,
\mathrm{km} \, \mathrm{s}^{-1}$ were considered (see e.g. Dressler et al.
\cite{dlb87}).

\begin{table*}
\caption[]{
Column 1    : cluster identification.
Column 2    : CMB-frame redshift.
Column 3    : limiting apparent magnitudes. The values were referred 
     to the distance of Coma by using the redshifts of column 2. 
Column 4    : number of galaxies (E+S0). 
Column 5    : references from which the FP parameters are drawn.
Column 6, 7 : FP slopes, $a$ and $b$, obtained by the \erlr fit, 
              with $1 \sigma$ standard intervals.
Column 8, 9 : FP slopes and relative uncertainites ($1 \sigma$ standard intervals) 
              as derived by the \erls method. 
Column 10   : intrinsic dispersion of the FP, $\mathrm{s}_{\log \sigma_0}$, 
              projected along the $\log \sigma_0$ direction. The values were 
              obtained by the \erls fit. 
     }
\centering
\begin{tabular}{|l|c|c|c|c|c|c|c|c|c|}
\hline
&&&&&  &  &  &  &\\
 Id. & $\mathrm{ c } z $ & $m_\mathrm{l}$ & 
  $N$ & Ref. & $a$  & $b$  & $a$  & $b$  & $\mathrm{s}_{\log \sigma_0}$ \\
& $(\mathrm{km \, s^{-1}})$ & & & & & & & & \\
&  &&&&  &  &  &  &\\
\hline
&&&&&  &  &  &  &\\
\ee &  6545 & 14.9 & 12 & JFK95a,b  &$1.14 \pm 0.20$&$0.306 \pm 0.033$ & $1.49 \pm 0.25$ & $0.300 \pm 0.043$  & 0.058 \\
\jj &  9800 & 13.8 & 13 & SLH97     &$0.77 \pm 0.24$&$0.317 \pm 0.022$ & $1.38 \pm 0.26$ & $0.330 \pm 0.040$  &  0.070 \\
\dd & 11472 & 14.8 & 14 & JFK95a,b &$1.10 \pm 0.13$&$0.249 \pm 0.028$ & $1.50 \pm 0.26$ & $0.232 \pm 0.026$  &  0.057 \\
\ff &  4421 & 14.7 & 16 & JFK95a,b &$0.89 \pm 0.16$&$0.341 \pm 0.020$ & $1.21 \pm 0.16$ & $0.353 \pm 0.021$  &  0.061 \\
\gg &  4050 & 15.1 & 18 & JFK95a,b &$1.19 \pm 0.14$&$0.320 \pm 0.028$ & $1.90 \pm 0.45$ & $0.316 \pm 0.033$  &  0.062 \\
\bb &  5038 & 14.7 & 21 & JFK95a,b &$0.94 \pm 0.12 $&$0.348 \pm 0.025$ & $1.26 \pm 0.13$ & $0.344 \pm 0.025$  &  0.058 \\
\ii &  4700 & 15.5 & 24 & SLH97     &$1.16 \pm 0.09$&$0.348 \pm 0.037$ & $1.52 \pm 0.31$ & $0.348 \pm 0.040$  &  0.056 \\
\cc &  8734 & 14.9 & 25 & JFK95a,b &$1.15 \pm 0.12$&$0.317 \pm 0.014$ & $1.51 \pm 0.23$ & $0.327 \pm 0.018$  &  0.029 \\
\hh &  4800 & 15.1 & 30 & SLH97     &$1.32 \pm 0.18$&$0.348 \pm 0.014$ & $1.68 \pm 0.17$ & $0.370 \pm 0.017$  &  0.010 \\
\aa &  7200 & 15.4 & 75 & JFK95a,b &$1.34 \pm 0.07$&$0.326 \pm 0.012$ & $1.53 \pm 0.08$ & $0.328 \pm 0.013$  &  0.015 \\
&&&&&  &  &  &  &\\
\hline
\end{tabular}
\label{cluster}
\end{table*} 

The FP coefficients were determined by the $\mathrm{MIST}_{\log r_\mathrm{e}},$ 
$\mathrm{MIST}_{\mu_\mathrm{e}},$~ \erls~ and \erbi~ fits (see Sect.~2.1), by 
taking into account the measurement errors on the variables as described 
in Sect.~2.2. 

\begin{figure} 
  \resizebox{\hsize}{7cm}{\includegraphics{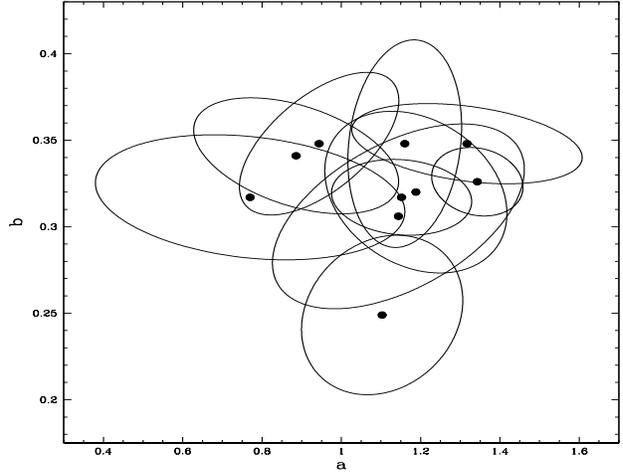}}
  \caption[]{ FP slopes, $a$ and $b$, of the samples listed in 
   Tab.~\ref{cluster}. The ellipses define a $10 \%$ CL. The 
   coefficients were derived by the \erlr~ fit. 
}
  \label{ellinc}
\end{figure}

To compare the FP slopes, it should be taken into account that 
the uncertainties on $a$ and $b$ can be correlated.
By Eqs.~B17, we obtained the theoretical estimates of the CM components
of the slopes. These estimates were then used to derive the ellipse that
defines a $10 \%$ CL for a normal bi-variate.
This is accounted for in Fig.~\ref{ellinc}. 

\begin{figure*}
  \resizebox{\hsize}{21cm}{\includegraphics{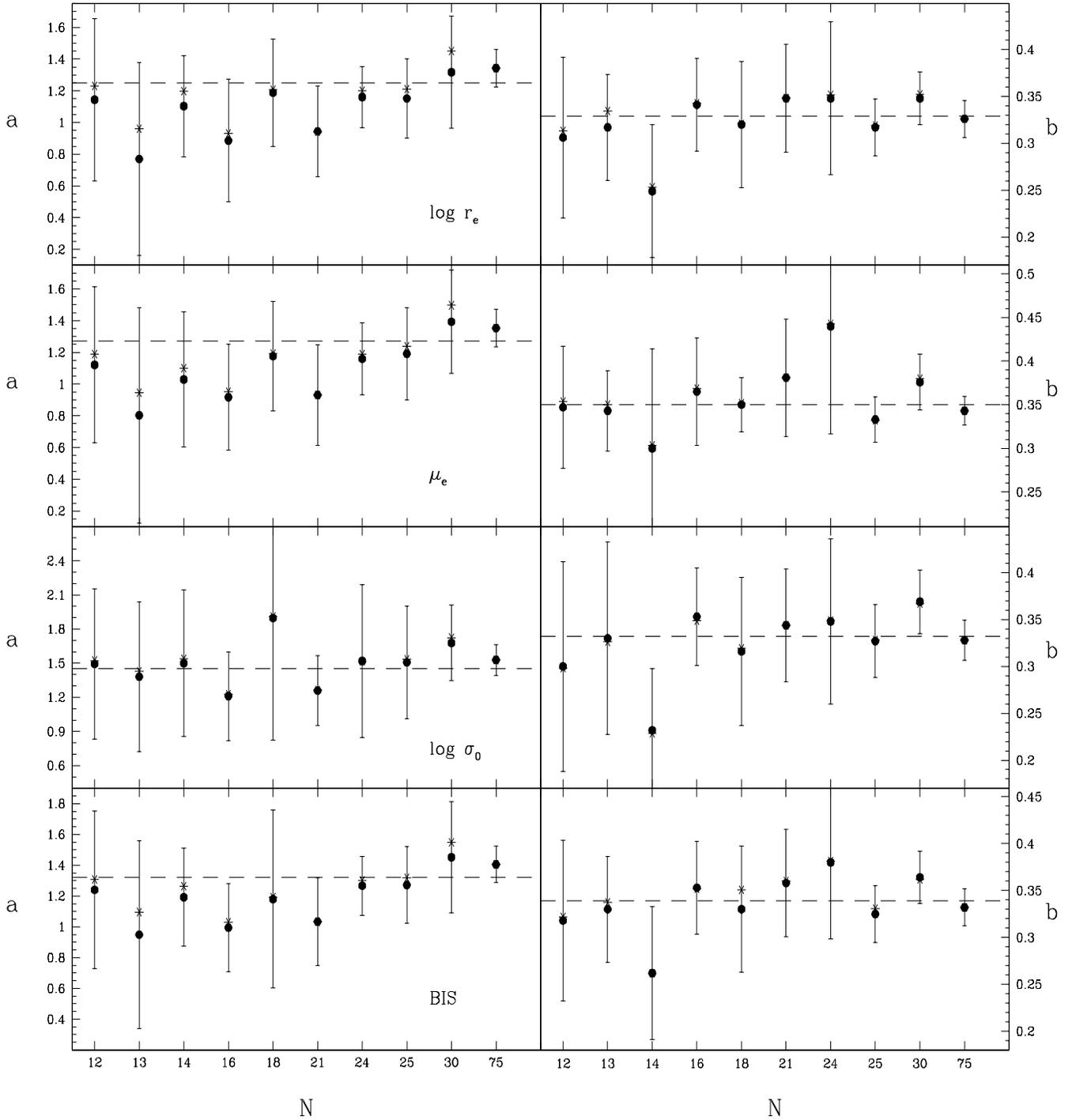}}
  \caption[]{FP slopes of the clusters listed in Tab.~\ref{cluster}.  
   The coefficients $a$ and $b$ of the different MIST fits are plotted
   on horizontal panels. The fitting methods are indicated shortly in the 
   lower right-hand corners of the panels on the left.
   The sample size increases from the left to the right of each panel
   as shown on the x-axes of the lower panels. The error 
   bars define a $10 \%$ CL. The values of $a$ and $b$ after the 
   magnitude-completeness corrections are also plotted by asterisks. 
   The dashed lines correspond to the weighted means of the slopes.
}
\label{cofcl}
\end{figure*}

The Fig.~\ref{ellinc} has the disadvantage to be not easily
readable. To obtain a more immediate description we compared 
separately the $a$ and $b$ values as derived from different fits, as 
plotted in Fig.~\ref{cofcl}.
The FP slopes of the \erlr~ and \erls~ fits are also listed 
in Tab.~\ref{cluster}. 

All the samples (except one) have small size, $N \le N_{\mathrm{min}} 
\approx 30$,  so that, as discussed in Sect.~3.2, an accurate estimate of 
the uncertainties on $a$ and $b$ is not achievable. 
Approximate error bars were derived by theoretic formulae (Eqs.~B17), with 
the prescriptions of Sect.~3.2.1 to obtain a $10 \%$ CL. Since half of the 
clusters have $N \widetilde{<} 20$, effective theoretic intervals should 
in fact be more suitable than re-sampling estimates (see Sect.~3.2.2).

Although the coefficients $a$ and $b$ are consistent for each pair 
of samples in every fitting procedure, we see that the error bars 
are very large, typically $40 \% - 50 \% $ of the $a$ and $b$ 
values. Moreover, they could not represent reliable estimates. 

It is also worth to notice that in the \erlr, \erbr~ and \erbi~ fits, 
the coefficient $a$ of the largest sample (the Coma cluster) is 
systematically higher with respect to the other determinations.
A weak correlation between $a$ and the sample size seems also
to be present.

To address this point, we tried to correct the FP slopes for the different
magnitude-completeness of the samples (see Giovanelli et al. \cite{ghh97}
and Scodeggio et al. \cite{sgb98} for a wide discussion). At first, we
constructed the completeness histograms of each sample. To this aim, the
magnitude range was binned and the fraction of galaxies of each bin
normalised to the corresponding fraction of the Coma photometric sample of
JFK95a. This data set is complete in fact out to a magnitude higher than
the magnitude limits of the other samples (see
Tab.~\ref{cluster} and references therein). It turns out that the same
results are also obtained by normalizing the fractions of galaxies through a
gaussian model of the luminosity function (see Scodeggio et al. 
\cite{sgb98}).
The histograms were then modeled by Fermi-Dirac distributions and
incomplete FP simulations constructed through the modeled distributions.
By fitting the simulations, we estimated the corrections on $a$ and $b$.

As shown in Fig.~\ref{cofcl}, the corrections shift upwards the 
coefficient $a$ and reduce the correlation with the sample size and 
the systematic difference of the Coma sample.
It turns out, in fact, that these effects are a consequence of the 
different completeness of the samples with respect to the photometric 
parameters. This is shown by the results of the `inverse' fit, with 
$\log \sigma_0$ as the dependent variable, that should be less sensitive 
to the photometric completeness (see Hudson et al. \cite{hls97}, hereafter
HLS97). 
By looking at Fig.~\ref{cofcl}, we see that in the \erls~ method there 
is no systematic difference between Coma and the other samples. 

It is also evident that a systematic difference is present between the
FP coefficients derived from the various MIST fits. As discussed in Sect.~2.2,
the `fitting bias' is due to the lack of a model for the intrinsic
scatter of the FP. For the same reason, only the projection of the
intrinsic dispersion along some assigned direction, with respect to a
given fitted plane, can be measured.

For instance, we calculate the scatter projection on the $\log \sigma_0$
variable for \erls~ fit.
For each sample we derived the \smr~ of the $\log \sigma_0$ residuals.
The amount of scatter due to the measurement errors was then subtracted in
quadrature. The measurement error scatter projected on $\sigma_0$ was
found to be $\sim 10 \%$. By constructing FP simulations with
$\sigma_0 \ge 100$, the `intrinsic \smr~ values' were corrected for
the bias due to the $\log \sigma_0$ cut on the cluster samples. The
correction was found to be very small ($\sim 2 $ -- $ 3 \%$) and largely
independent of the simulation parameters.

The $\log \sigma_0$ dispersions are shown in Tab.~\ref{cluster}. The mean
values of the $\log \sigma_0$ intrinsic scatter amounts to $\sim
0.048$, that corresponds to $\sim 10 \%$ in $\sigma_0$.

In Tab.~\ref{means}, we show the weighted means of the coefficients $a$
and $b$ for the various fitting methods. The means were calculated
after the magnitude-completeness and the $\log \sigma_0$ 
corrections. The $\log \sigma_0$ bias on the slopes of the \erls~ fit
was estimated from the simulations (see above). It amounts to $\sim 5
\%$ for $a$ and is completely negligible for $b$.

The difference between the various MIST fits amounts to $\sim 15 \%$
for the coefficient $a$ and to $\sim 6 \%$ for $b$. The coefficient
$a$ varies from $1.25$ to $1.45$, and $b$ from $0.33$ to $0.35$.

The bias that could be introduced by neglecting the measurement errors was
found to be negligible for $b$ ($\widetilde{<} 1 \%$). Concerning
the coefficient $a$, the bias depends on the fitting procedure: it
amounts to $\sim 5 \%$ for the \erlr~ and \erbr methods (in agreement
with the estimate given by JFK96), and to $\sim 1 \%$ and $\sim 3 $ --
$ 4 \%$ for the \erls~ and \erbi~ fits. Because of the correlation
between the uncertainties on \lr~ and \br, it turns out that the \erls~
method is quite insensitive to the uncertainties on the FP parameters.

For what concerns the covariance term of the photometric parameters, as
discussed in Sect.~2.2, it was estimated by the relation
$\mathrm{C}(\log r_\mathrm{e}, \mu_\mathrm{e}) \approx \alpha
\cdot \mathrm{V}( \mu_e )$ with $\alpha \approx 0.30$ (see JFK95a). 
By analyzing the correlation of the errors on $\log
r_\mathrm{e}$ and $\mu_\mathrm{e}$ for various photometric data
set drawn from literature, we obtained values of $\alpha$ included
between $0.26$ and $0.34$. Varying $\alpha$ in this range for
the calculation of $\mathrm{C}(\log r_\mathrm{e},
\mu_\mathrm{e})$, very small variations ($< 1 \%$) were obtained
in the FP coefficients. The higher systematic difference, 
$\sim 2$ -- $3 \%$, was found for the coefficient $a$ in the 
\ls~ fit.

\begin{figure}
  \resizebox{\hsize}{10cm}{\includegraphics{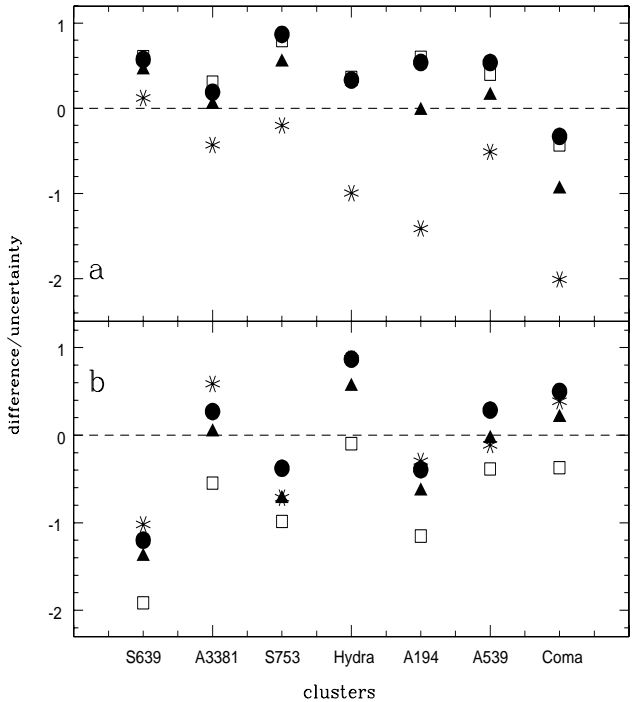}}
  \caption[]{ Comparison of the FP slopes obtained by the MIST fits with 
   the values of JFK96 (see text). Symbols for the different
   MIST fits are as in Fig.~\ref{disp}.
}
\label{confronto}
\end{figure}

\begin{table}
\caption[]{ Weighted means of the 'corrected' FP slopes (see text). 
  Column 1: fitting method.
  Column 2 and 3: mean values of $a$ and $b$ with corresponding 
  uncertainties (1$\sigma$ intervals). }
\centering
\begin{tabular}{|l|c|c|}
\hline
&&\\
 & $ \overline{a} $ &
   $ \overline{b} $ \\
&&\\
\hline
&&\\
 \erlr & $1.25 \pm 0.05$ & $ 0.329 \pm 0.008$ \\
 \erbr & $1.27 \pm 0.05$ & $ 0.350 \pm 0.007$ \\
 \erls & $1.45 \pm 0.06$ & $ 0.332 \pm 0.009$ \\
 \erbi & $1.32 \pm 0.04$ & $ 0.339 \pm 0.007$ \\
&&\\
\hline
\end{tabular}
\label{means}
\end{table}

The systematic dependence of the FP coefficients from the fitting method 
can also be seen by comparing the values obtained by the MIST fits with the 
results of JFK96 and HLS97.
 
In Fig.~\ref{confronto} the MIST estimates of $a$ and $b$ are compared
to the values of JFK96, that adopted an ORLS (robust) method. To this aim, 
we plot the differences of the FP slopes (JFK96 $-$ ours) divided 
by the relative $1 \sigma$ uncertaintities added in quadrature. The values 
obtained by JFK96 appear to be consistent with our estimates for each 
cluster. However, some systematic difference exists. The values of $a$ obtained 
by the \erlr~ and \erbr~ fits are systematically lower than those of JFK96. 
A small systematic difference is also found for the \erbi~ method. For 
the \ls~ fit the MIST estimates are systematically higher.  \\   
The values of $b$ agree with those by JFK96 for the \erlr, \erbr~ and 
the \erbi~ methods. On the other hand, the values of the \erbr~ fit are 
systematically higher. \\ 
The values of the FP slopes obtained by JFK96 for the whole cluster sample, 
$a = 1.24 \pm 0.07$ and $b = 0.328 \pm 0.008$, are generally consistent 
with the mean values reported in Tab.\ref{means}. However, systematic 
differences are particularly evident for the value of $a$ in the \erls~ 
fit and for the value of $b$ in  the \erbr~ method.       

HLS97 derive the FP slopes for a sample of seven clusters of galaxies. 
By adopting the $\mathrm{OLS}_{\log \sigma_0}$ method on all the clusters
simultaneously, they obtain $a = 1.383 \pm 0.04$ and $b = 0.326 \pm 0.011$.
These values are consistent with the mean values of the \erls~ fit 
(see Tab.~\ref{means}). \\
The slopes of the individual clusters of HLS97 are derived by a bi-dimensional
fit, adopting as independent variable the combination of \lr~ and \br~
obtained from the global fit. A direct comparison of our estimates with
the individual results of HLS97 is thus not possible.

Since the ORLS and OLS methods do not account for the measurement errors, 
and due to differences in completeness and selection, a detailed 
explanation of the origin of the above discrepancies is not achievable.

\section{Summary and conclusions.}

The FP of cluster galaxies is one of the most promising tools to study
galaxy evolution, to constrain the epoch of galaxy formation, and to set
constraints on the geometry of the universe. To those aims the FP has
been recently studied in intermediate-redshift ($z=0.18 - 0.83$) clusters
(van Dokkum \& Franx \cite{vdf96}, Kelson et al. \cite{kivd97}, Bender et
al. \cite{bsz98}, van Dokkum et al. \cite{vdf98}, J\o rgensen et al.
\cite{jor99}, Kelson et al. \cite{kivd00}). The main step in the above
applications is the comparison between FP determinations in different
clusters, wavebands and redshifts, for which the determination of the FP
coefficients and of the relative uncertainties plays a crucial r\^ole.

The present study aims at clarifying various aspects in the
determination of the FP coefficients and uncertainties.

\subsection{The problem of fitting the FP.}

We introduced a statistical model of the FP that takes into account
the measurement errors on the variables and the intrinsic scatter. The
correlation between measurement uncertainties is also accounted for.

We derived the MIST$_i$ (\emph{Measurement errors and Intrinsic
Scatter Three dimensional}) fitting procedures, where $i = \log
r_{\mathrm{e}}, \mu_{\mathrm{e}}, \log \sigma_0$ is the dependent
variable of the fit. A bisector method, \erbi, is also introduced.

The assumptions under which the MIST procedures give unbiased
estimates of the FP coefficients are that the errors and the intrinsic
scatter do not depend on the `location' on the fitting plane (H1),
their  covariance matrices are known (H2 and H3), and the covariance
matrices of the observed quantities are known (H4).

Under these assumptions, the fits give the same coefficients whatever
the choice of the dependent variable.

For what concerns the above assumptions, the knowledge of the average
measurement errors can be easily introduced in the procedures in order
to satisfy H1 and H2, while the problems posed by H4 can be overcome
by assuming a normal distribution for the observed quantities.

Assumption H3, on the other hand, cannot be satisfied since we do not
have a model for the intrinsic dispersion of the FP variables. We
showed that it is precisely the lack of such a model that biases 
the best-fit coefficients in such a way that their values depend 
on the choice of the dependent variable and could not coincide with
the slopes of the `true relation'.

\subsection{Uncertainties on the FP coefficients.}

We addressed the problem of the estimate of the uncertainties on the
FP coefficients by numerical simulations based on the best sampled FP
so far (the Gunn $r$ FP of Coma by JFK96).

The results of the simulations are presented in Fig. 4 in terms of
$\delta a/a$ and $\delta b/b$ versus $ \log( N )$. 
This figure can be used to state the sample size needed to achieve a
given accuracy for the FP coefficients.

On the basis of the simulations we tested the performances of the
methods used to estimate the uncertainties: the theoretical formulae
and the bootstrap technique.

We showed that for $N > N_{min} \approx 30$ the theoretical
formulae can recover the variances of the FP coefficients within about
5\%, while for $N < N_{min}$ the uncertainties are systematic smaller than
the true values and, what is more troublesome, are affected by a large
scatter.

The bootstrap technique gives, on the average, more accurate estimates
than theoretical methods but for small samples the scatter of the
individual values is much larger. 
Again, a minimum sample size of $N = N_{min} \approx 30$ exists for the 
bootstrap to give reliable estimates.

\subsection{Implications for the use of the FP.}

As an application of the present study we addressed in Sect. 4
the question of the `universality' of the FP.
The main obstacles are the limited size of the available (homogeneous)
samples and their different completeness characteristics.

We find that the slopes of the FP in the ten clusters considered
are consistent \emph{ within the large uncertainties} (due to the
small number of galaxies in most of the samples). We stress that
this result should be taken at most as an indication in favor of the
universality of the FP and not as an evidence.

With the available data samples it is not possible to settle the
question of the universality: to this aim, larger and more
homogeneous samples would be needed.

For the application of the FP to the study of galaxy evolution it would
be crucial to know the behavior of the slopes as a function of redshift.
Pahre et al. (1998) built a model for the FP that predicts a change in
slope with redshift: if age contributes to the tilt of the FP, massive
galaxies should be much older than the least luminous ones. As a
consequence of this `differential' evolution, the slopes will change with
redshift. They predict the change with redshift of the coefficient $a$ of
the FP (see their Fig. 10) by adopting two different models for the
stellar evolution (Bruzual \& Charlot \cite{brc98} and Vazdekis et al.
\cite{vcp96}). The Bruzual \& Charlot models produces a rapid change of
$a$, that should decrease by about 0.15 from z=0 to z$\sim$0.3, namely,
12\% of its local value in the Gunn $\mathrm{r}$ band.

Recently, Kelson et al. (\cite{kivd00}) derived the FP for a cluster at
z=0.33 with 30 galaxies. The authors claim that the slopes are
fully consistent with those of the local FP (Coma). They derive the
FP by the orthogonal fit and give the following uncertainties for
the slopes: $\delta a/a$=0.1 and $\delta b/b$=0.11, that are 
consistent with the values obtained here from the simulations with 
$n$=30 (see Fig. 4). These data seem thus to rule out the rapid
evolution of $a$ predicted with the Bruzual \& Charlot model.

The situation is quite different with the Vazdekis et al. models. At
$z=0.3$ the slope $a$ decreases by only 0.03\%. The minimum sample size
needed to achieve such a precision at $z=0.3$ would be $N \approx 300$,
which is out of the reach of any possible observation. The situation
improves by moving at higher redshifts, where the magnitude of the
predicted change of the slopes is larger. For the cluster at z=0.83
studied by van Dokkum et al. (\cite{vdf98}), in fact, the situation is by far
more promising. To test the predicted change in slope it would be
sufficient to determine the FP on $N \approx 40$ galaxies, that is well
within the possibility of a 8-10m class telescope in a reasonable number
of nights.

\appendix{ }
\section*{\bf Appendix A}

From Eq.~(\ref{relsca}), we write the probability density (hereafter PD) of 
$Y_i$ as:
\begin{eqnarray*}
  f_{ Y_i}( Y_i ) d \! Y_i =     
 \int f_{ X_i }( X_i) \cdot 
  { f_{\Phi_i} }( Y_i - X_i ; X_i) d \! X_i    
  \hspace{1.3cm} (\mathrm{A}1) 
\end{eqnarray*}
where $f_{Y_i}(y_i) $ and $f_{X_i}(x_i)$ are the PDs of $Y_i$  
and $X_i$, and ${ f_{\Phi_i} }( \phi_i ; X_i) d \! \phi_i   $ 
is the probability that, for a fixed output of $X_i$, $\Phi_i $ 
is included between $\phi_i$ and $\phi_i + d \! \phi_i$. For brevity, 
we put ${\Phi_\mathrm{m}}_i +{\Phi_\mathrm{s}}_i = {\Phi_i}$.

By Eq.~(A1) we obtain the following relations:
\begin{eqnarray*}
\mathrm{E}( Y_i ) 
 & = & \int Y_i f_{Y_i}(Y_i) d \! Y_i = \\
 & = & \int f_{ X_i}( X_i)  
       \left[ \int Y_i { f_{\Phi_i} }( Y_i - X_i ; 
       X_i) d \! Y_i \right] d \! X_i =  \\
 & = & \int X_i f_{ X_i}( X_i)  
   \left[ \int   { f_{\Phi_i} }(\phi_i ; X_i) d \! \phi_i \right] 
    d \! X_i + \\ 
 & + & \int f_{X_i}(X_i)  \left[ 
    \int \phi_i { f_{\Phi_i} }(\phi_i ; X_i) d \! \phi_i \right] 
    d \! X_i \hspace{1.1cm} (\mathrm{A}2)
\end{eqnarray*}
\begin{eqnarray*}
 \mathrm{V}( Y_i )  
  & = & \int \left( Y_i^2 - \mathrm{E}(Y_i)^2 \right) 
        f_{Y_i}(y_i) d \! y_i = \\ 
  & \hspace{-1.4cm} = & \hspace{-0.7cm} 
       \int \left[ x_i^2 - {\mathrm{E}(X_i)}^2 \right] 
        f_{ X_i}(X_i)  \left[ \int   
        { f_{\Phi_i} }(\phi_i ; X_i) d \! \phi_i \right] 
        d \! X_i  +  \\ 
  & \hspace{-1.4cm} + & \hspace{-0.7cm} \! 2 \int X_i f_{X_i}(X_i)  
        \left[ \int \phi_i  { f_{\Phi_i} }(\phi_i ; X_i) 
        d \! \phi_i \right] d \! X_i \\
  & \hspace{-1.4cm} + & \hspace{-0.7cm} \! \int f_{ X_i}(X_i) 
        \left[ \int \phi_i^2  
        f_{\Phi_i}(\phi_i ; X_i) d \! \phi_i \right] 
        d \! X_i  \hspace{1.8cm} (\mathrm{A}3)
\end{eqnarray*}

By using the normalization of $f_{\Phi_i}$ and the hypothesis 
$\mathrm{E}( \Phi_i ) = 0$, from (A2) we obtain Eq.~(\ref{glosa}).
Eq.~(\ref{glosb}) follows from (A3) with the further hypothesis that
$\Phi_i$ and \xxi~ are not correlated ('hypothesis \hyp').

If \hyp~ does not hold, we obtain again Eq.~(\ref{glosb}) provided that 
the CM terms of $ \left\{ {\Phi_\mathrm{m}}_i \right\}$ and 
$ \left\{ {\Phi_\mathrm{s}}_i \right\}$ are substituted with their expected 
values with respect to the PDs of \xxi.

In a similar way, Eq.~(\ref{glosc}) can be proved.

\appendix{ }
\section*{\bf Appendix B}

Let's start by introducing the following notations:
\begin{eqnarray*}
  \mathrm{V}_{ij} & = & \mathrm{C}( \Phi_{\mathrm{m}_i} ,
     \Phi_{\mathrm{m}_j} ) + \mathrm{C}( \Phi_{\mathrm{s}_i} ,
     \Phi_{\mathrm{s}_j} ) \hspace{2.92cm}  (\mathrm{B}1)\\
  \mathrm{C}_{ij} & = & \mathrm{C}( Y_i , Y_j ) 
                    \hspace{5.63cm} (\mathrm{B}2) \\
  \Delta   & = & \left[ \mathrm{C}_{12} - \mathrm{V}_{12} \right]^2 -
                 \left[ \mathrm{C}_{11} - \mathrm{V}_{11} \right] 
                 \left[ \mathrm{C}_{22} - \mathrm{V}_{22} \right]   
                \hspace{1.07cm} (\mathrm{B}3) \\
  \Delta_1 & = & \left[ \mathrm{C}_{23} - \mathrm{V}_{23} \right]  
                 \left[ \mathrm{C}_{12} - \mathrm{V}_{12} \right] 
               - \left[ \mathrm{C}_{13} - \mathrm{V}_{13} \right] 
               \left[ \mathrm{C}_{22} - \mathrm{V}_{22} \right]  \\  
           & &    \hspace{7.0cm} (\mathrm{B}4) \\
  \Delta_2 & = & \left[ \mathrm{C}_{12} - \mathrm{V}_{12} \right]  
                  \left[ \mathrm{C}_{13} - \mathrm{V}_{13} \right] 
                 -  \left[ \mathrm{C}_{23} - \mathrm{V}_{23} \right]   
                  \left[ \mathrm{C}_{11} - \mathrm{V}_{11} \right] \\  
           &   &   \hspace{7.0cm} (\mathrm{B}5) 
\end{eqnarray*}
The symbols $\hat{ \Delta }$, $\hat{ \Delta_1 }$ and $\hat{ \Delta_2 }$ 
will indicate the quantities obtained by substituting $\mathrm{C}_{ij}$ 
in Eqs.~(B3 -- B5) with unbiased estimates $\hat{ \mathrm{C} }_{ij} $.

From Eqs.~(\ref{fita} -- \ref{fitd}) we have:

\begin{eqnarray*}
 \hat{ \alpha } - \alpha & = & \left\{ \frac{ \hat{ \Delta_1 } }{ \hat{ \Delta } } - 
        \frac{\Delta_1}{\Delta} \right\} = \\
   & = & \frac{1}{\Delta}  
        \left\{ \left( \hat{ \Delta_1 } - \Delta_1 \right) - 
        \alpha \left( \hat{ \Delta } - \Delta \right) 
        \right\} + o \hspace{1.25cm} (\mathrm{B}6) \\
 \hat{ \beta } - \beta & = & 
                     \frac{1}{\Delta} \left\{ \left( \hat{ \Delta_2 } - 
                     \Delta_2 \right) - \beta  
                     \left( \hat{ \Delta } - \Delta \right)  \right\} 
	                 + o \hspace{1.25cm} (\mathrm{B}7)
\end{eqnarray*}
where $\hat{ \alpha }$ and $\hat{ \beta }$ are defined by replacing 
${ \mathrm{C} }_{ij} $ with $\hat{ \mathrm{C} }_{ij} $ in 
Eqs.~(\ref{fita} , \ref{fitb}), and $o$ is a quantity 
that can be neglected when the sample size increases.

By using the relation
\begin{eqnarray*}
  \hat{ \mathrm{E} }( {\Sigma} ) = \mathrm{E}( \Sigma )  
  + o \hspace{5.5cm} (\mathrm{B}8)
\end{eqnarray*}
 where $\Sigma$ is a generic RV and $\hat{ \mathrm{E} }( {\Sigma} )$
  an unbiased estimate of $\mathrm{E}( {\Sigma} )$, the following 
relations can be written: 
\begin{eqnarray*}
\hat{ \mathrm{C} }_{ij} - \mathrm{C}_{ij} & = & 
      \hat{ \mathrm{E} }(Y_i Y_j) - 
      \mathrm{E}(Y_i Y_j) 
       - \hat{ \mathrm{E} }( Y_i) \mathrm{E}( Y_j) + \\
    & - & \mathrm{E}( Y_i) \hat{ \mathrm{E} }( Y_j ) + 
          o \hspace{2.7cm} (\mathrm{B}9) \\
  \hat{ \mathrm{C} }_{ij} 
   \hat{ \mathrm{C} }_{kl} - \mathrm{C}_{ij} \mathrm{C}_{kl} 
  & = & \mathrm{C}_{ij}  
     \left\{ \left[ \hat{ \mathrm{E} }( Y_k 
     Y_l ) -
     \mathrm{E}( Y_k Y_l) \right] \right. + \\ 
  & \hspace{-4.8cm} + & \hspace{-2.5cm} \left.
     \mathrm{E}( Y_k) \left[ \hat{\mathrm{E}}( Y_l) - 
     \mathrm{E}( Y_l)  \right] - \mathrm{E}(Y_l)   
     \left[ \hat{ \mathrm{E} }(Y_k) - \mathrm{E}(Y_k) 
     \right] \right\} + \\ 
  & \hspace{-4.8cm} + & \hspace{-2.5cm} \mathrm{C}_{kl}  
         \left\{ \left[ \hat{ \mathrm{E} }( Y_i 
         Y_j) - \mathrm{E}( Y_i Y_j) \right] -    
         \mathrm{E}( Y_i) \left[ \hat{ \mathrm{E} }(Y_j) 
         - \mathrm{E}( Y_j) \right] \right. + \\
 & \hspace{-4.8cm} - & \hspace{-2.5cm} \left. 
         \mathrm{E}( Y_j) \left[ \hat{ \mathrm{E} }(Y_i) 
         - \mathrm{E}( Y_i) \right] \right\} \! + o 
     \hspace{3.25cm} (\mathrm{B}10) 
\end{eqnarray*}
Setting 
$\mathrm{P}_{ijlk} = \left( \mathrm{C}_{ij} - \mathrm{V}_{ij} \right)
Y_l Y_k $
and 
$\mathrm{q}_{ijl} = \left( \mathrm{C}_{ij} - \mathrm{V}_{ij} \right)  
\mathrm{E}( Y_l )$, we introduce the RVs $ \zeta $, $ \zeta_1 $ and 
$ \zeta_2 $ by the following definitions: 
\begin{eqnarray*}
\zeta \cdot \Delta & = & 2 \mathrm{P}_{1212} - \mathrm{P}_{1122} - 
          \mathrm{P}_{2211} + 2 \left[ \mathrm{q}_{221} - 
          \mathrm{q}_{122} \right] \mathrm{Y_1} + \\
 & \hspace{-1cm} + & \hspace{-0.5cm}
          2 \left[ \mathrm{q}_{112} - \mathrm{q}_{121} \right] 
            \mathrm{Y_2} \hspace{4.4cm} (\mathrm{B}11) \\
\zeta_1 \cdot \Delta & = & \mathrm{P}_{2312} + \mathrm{P}_{1223} -
          \mathrm{P}_{2213} - \mathrm{P}_{1322} + \left[  
           2 \mathrm{q}_{132} - \mathrm{q}_{123} + \right. \\
 & \hspace{-1cm} - & \hspace{-0.5cm} \left. \mathrm{q}_{231}  
         \right] \mathrm{Y_2} +  
           \left[ \mathrm{q}_{223} - \mathrm{q}_{232} 
           \right] \mathrm{Y_1} +  \left[ \mathrm{q}_{221} - 
           \mathrm{q}_{122} \right] \mathrm{Y_3} 
           \hspace{0.19cm} (\mathrm{B}12) \\  
\zeta_2 \cdot \Delta & = & \mathrm{P}_{1312} + \mathrm{P}_{1213} - 
           \mathrm{P}_{1123} - \mathrm{P}_{2311} + \left[ 2  
           \mathrm{q}_{231} - \mathrm{q}_{123} + \right. \\
 & \hspace{-1cm} - & \hspace{-0.5cm} \left.  \mathrm{q}_{132} \right] 
  \mathrm{Y_1} 
   + \left[ \mathrm{q}_{113} - \mathrm{q}_{131} \right] \mathrm{Y_2} +
          \left[ \mathrm{q}_{112} - \mathrm{q}_{121} \right] \mathrm{Y_3}  
         \hspace{0.19cm}  (\mathrm{B}13)
\end{eqnarray*}
Using Eqs.~(B8 -- B13) and setting 
$\zeta_{\alpha} = \zeta_1 - \alpha \zeta$, 
$\zeta_{\beta} = \zeta_2 - \beta \zeta$, after lengthy calculations, 
Eqs.~(B6,B7) are rewritten:
\begin{eqnarray*}
  \hat{ \alpha } - \alpha & = & \hat{ \mathrm{E} }( \mathrm{ \zeta_\alpha } ) -  \mathrm{E}( \zeta_\alpha  )  + o 
            \hspace{3.69cm} (\mathrm{B}14) \\
  \hat{ \beta } - \beta & = & 
          \hat{ \mathrm{E} }( \zeta_\beta ) - \mathrm{E}( \zeta_\beta ) 
         + o \hspace{3.7cm} (\mathrm{B}15)
\end{eqnarray*}
From the above relations and Eq.~(\ref{fitc}), we also have:
\begin{eqnarray*}
\hat{ \gamma } - \gamma & = & 
    \hat{\mathrm{E}}( \mathrm{\zeta_\gamma} ) - 
    \mathrm{E}( \mathrm{ \zeta_\gamma } ) + 
     o \hspace{3.78cm} (\mathrm{B}16)
\end{eqnarray*}
where $ \mathrm{\zeta_\gamma} = Y_3 - \alpha Y_1 - 
  \beta Y_2 - \mathrm{E}( Y_1) \zeta_\alpha -
  \mathrm{ E }( Y_2) \zeta_\beta $. 
  
If \yi~ are normally distributed, estimates of the expected values 
in the previous equations are given by sample means. 
The CM terms of $ { \alpha }$, ${ \beta }$ and $ { \gamma }$ are 
thus obtained by the usual formulae:
\begin{eqnarray*} \hat{ \mathrm{C} }( \hat{ \mu } , \hat{ \nu } ) & = & \frac{1}{N (N-1)} 
   \sum_{i=1,N} ( \hat{\zeta_{\mu}}_i - \overline{\hat{\zeta_{\mu}}}) 
   ( \hat{\zeta_{\nu}}_i - \overline{\hat{\zeta_{\nu}}}) \\
 & & \hspace{3.765cm}
   \mu, \nu = { \alpha }, { \beta }, { \gamma } 
   \hspace{0.5cm} (\mathrm{B}17)
\end{eqnarray*}
where $N$ is the sample size and $\hat{\zeta_{\alpha}}$,
 $\hat{\zeta_{\beta}}$ and $\hat{\zeta_{\gamma}}$ are the RVs obtained 
 by substituting in the expressions of $\zeta_{\alpha}$, $\zeta_{\beta}$ 
 and $\zeta_{\gamma}$ the unknown quantities with their unbiased
 estimates.

\begin{acknowledgements}
We thank the referee for carefully reading the manuscript and 
for helpful suggestions.
\end{acknowledgements}


\begin{thebibliography}{}
\bibitem[1996]{akb96} Akritas, M.G., Bershady, M.A., 1996, 
                         ApJ, 470, 706  
\bibitem[1990]{bfg90} Beers, T.C., Flynn, K., Gebhardt, K., 1990,
                         AJ, 100, 32
\bibitem[1992]{bbf92} Bender, R., Burstein, D., Faber, S.M., 1992,
                         ApJ, 399, 462
\bibitem[1998]{bsz98} Bender, R., Saglia, R.P., Ziegler, B.,  
                         et al., 1998, ApJ, 493, 529
\bibitem[1998]{brc98} Bruzual, G.A., Charlot, S., 1998, ApJ  
                         in preparation  
\bibitem[1997]{bcc97} Busarello, G., Capaccioli, M., Capozziello, S., 
                         et al., 1997, ApJ, 320, 415
\bibitem[1996]{clr96} Ciotti, L., Lanzoni, B.,  Renzini, A., 
                         1996, MNRAS, 282, 1 
\bibitem[1992]{dcd92} de Carvalho, R.R., Djorgovski, S., 
                         1992, ApJL, 389, L49
\bibitem[1987]{djd87} Djorgovski, S., Davies, M., 
                         1987, ApJ, 313, 59
\bibitem[1997]{dcz97} D'Onofrio, M., Capaccioli, M., Zaggia, S., et al.,
                         1997, MNRAS, 289, 847                       
\bibitem[1987]{dlb87} Dressler, A., Lynden-Bell, D., Burstein, D., 
                         et al., 1987, ApJ, 313, 42
\bibitem[1979]{efr79} Efron, B., 1979, Ann. Statist., 7, 1 
\bibitem[1986]{eft86} Efron, B., Tibshirani, R.J., 1986, 
                         Statist. Science, 1, 54 
\bibitem[1987]{efr87} Efron, B., 1987, J. Amer. Statist. Assoc., 
                      82, 171 
\bibitem[1987]{fdd87} Faber, S.M., Dressler, A., Davies, L.R., et al., 1987,
                      in: Nearly Normal Galaxies,
                      ed.\ S.M. Faber, Springer Verlag, 
                      New york, p.\ 175
\bibitem[1992]{feb92} Feigelson, E.D.,  Babu, G.J., 
                         1992, ApJ, 397, 55  
\bibitem[1997]{ghh97} Giovanelli, R., Haynes, M.P., Herter, T., et al., 
                         1997, AJ, 113, 22
\bibitem[1996]{grc96} Graham, A.W.,  Colless, M., 1996, 
                         MNRAS, 287, 221
\bibitem[1998]{gra98} Graham, A.W., 1998, MNRAS, 295, 933
\bibitem[1997]{hls97} Hudson, M.J., Lucey, J.R., Smith, R.J., et al., 1997,
                         MNRAS, 291, 488
\bibitem[1990]{ifa90} Isobe, T., Feigelson, E.D., Akritas, M.G.,  et al., 
                         1990, ApJ, 364, 104
\bibitem[1995a]{jfk95a} J\o rgensen, I., Franx, M.,  Kj\ae rgaard, P., 
                         1995(a), MNRAS, 273, 1097                              
\bibitem[1995b]{jfk95b} J\o rgensen, I., Franx, M.,  Kj\ae rgaard, P., 
                         1995(b), MNRAS, 276, 1341
\bibitem[1996]{jfk96} J\o rgensen, I., Franx, M.,  Kj\ae rgaard, P., 
                         1996, MNRAS, 280, 167
\bibitem[1999]{jor99} J\o rgensen, I., Franx, M., Hjorth, J., et al.,  
                         1999, MNRAS, 308, 833
\bibitem[1993]{kjm93} Kj\ae rgaard, P., J\o rgensen, I., Moles, M., 
                         1993, ApJ, 418, 617        
\bibitem[1997]{kivd97} Kelson, D.D., van Dokkum, P.G., Franx, M.,  
                         et al., 1997, ApJ, 478, L13                  
\bibitem[2000]{kivd00} Kelson, D.D., Illingworth, G.D., van Dokkum, P.G., 
                         et al., 2000, ApJ, 531, 184
\bibitem[1991]{lbe91} Lucey, J.R., Bower, R.G.,  Ellis, R.S., 
                         1991, MNRAS, 249, 755  
\bibitem[1999]{mga99} Mobasher, B., Guzman, R., Aragon-Salamanca, A., et al.,
                         1999, MNRAS, 304, 225
\bibitem[1995]{pdc95} Pahre, M.A., Djorgovski, S.G., de Carvalho, R.R.,
                         1995, ApJ, 453, L17
\bibitem[1998]{pcd98} Pahre, M.A., de Carvalho, R.R.,  Djorgovski, S.G.,
                         1998, AJ, 116, 1606
\bibitem[1994]{prs94} Prugniel, P.,  Simien, F., 1994, A\&A, 282, L1
\bibitem[1996]{prs96} Prugniel, P.,  Simien, F., 1996, A\&A, 309, 749
\bibitem[1997]{prs97} Prugniel, P.,  Simien, F., 1997, A\&A, 321, 111
\bibitem[1949]{que49} Quenouille, M., 1949, J. R. Statist. Soc. B, 11, 18
\bibitem[1990]{rcs90} Recillas-Cruz, E., Carrasco, L., Serrano, P.G., et al.,
                         1990, A\&A, 229, 64
\bibitem[1991]{rcs91} Recillas-Cruz, E., Carrasco, L., Serrano, P.G., et al.,
                         1991, A\&A, 249, 312                         
\bibitem[1993]{rec93} Renzini, A., Ciotti, L., 1993, ApJ, 416, L49
\bibitem[1998]{sgb98} Scodeggio, M., Gavazzi, G., Belsole, E., et al.,
                         1998, MNRAS, 301, 1001
\bibitem[1997]{slh97} Smith, R.J., Lucey, J.R., Hudson, M.J., et al.,
                         1997, MNRAS, 291, 461
\bibitem[1958]{tuk58} Tukey, J., 1958, Ann. Math. Statist., 29, 614
\bibitem[1996]{vdf96} van Dokkum, P.G., Franx, M., 1996, MNRAS, 281, 985
\bibitem[1998]{vdf98} van Dokkum, P.G., Franx, M., Kelson, D.D.,
                         et al., 1998, ApJ, 504, L17
\bibitem[1996]{vcp96} Vazdekis, A., Casuso, E., Peletier, R., et al.,  
                         1996, ApJS, 106, 307 
\bibitem[1997]{zib97} Ziegler, B.L., Bender, R., 
                         1997, MNRAS, 291, 527
\bibitem[1999]{zsb99} Ziegler, B.L., Saglia, R.P., Bender, R., et al., 
                         1999, A\&A, 346, 13
\end{thebibliography}
\end{document}